\documentclass[fleqn,usenatbib]{mnras}

\usepackage{newtxtext,newtxmath}
\usepackage[T1]{fontenc}

\DeclareRobustCommand{\VAN}[3]{#2}
\let\VANthebibliography\thebibliography
\def\thebibliography{\DeclareRobustCommand{\VAN}[3]{##3}\VANthebibliography}

\usepackage{graphicx}	% Including figure files
\usepackage{amsmath}	% Advanced maths commands
\usepackage{amssymb}	% Extra maths symbols

\title[M92: revisiting its variable star population]{CCD \emph{VI} time-series of the extremely metal poor globular cluster M92: revisiting its variable star population$\thanks{Based on observations made with 0.84m and 1.5m telescopes of the San Pedro M\'artir Observatory, Mexico.}$}

\author[Yepez et al.]{
M. A. Yepez,$^{1}$\thanks{E-mail: myepez@astro.unam.mx}
A. Arellano Ferro,$^{1}$
D. Deras$^{1}$
\\
$^{1}$Instituto de Astronom\'ia, Universidad Nacional Aut\'onoma de M\'exico, Ciudad de M\'exico, CP 04510, M\'exico.\\
}

\date{Accepted 2020 February 27. Received 2020 February 26; in original form 2020 February 11}

\pubyear{2020}

\begin{document}
\label{firstpage}
\pagerange{\pageref{firstpage}--\pageref{lastpage}}
\maketitle

\begin{abstract}
We present an analysis of \emph{VI} CCD time-series photometry of the Oo II type globular cluster M92. The variable star population of the cluster is studied with the aim of revising their classifications, identifications, frequency spectra and to select indicators of the parental cluster metallicity and distance. The Fourier decomposition of RR Lyrae light curves lead to the estimation of mean [Fe/H]$_{\rm spec}=-2.20\pm 0.18$, and distance of $8.3\pm0.2$ kpc. Four new variables are reported; one RRd (V40), a multimode SX Phe (V41), a SR (V42) and one RRc (F1) that is most likely not a cluster member. The AC nature of V7 is confirmed. The double mode nature of the RRc star V11 is not confirmed and its amplitude modulations are most likely due to the Blazhko effect. Two modes are found in the known RRc variable V13. It is argued that the variable V30, previously classified as RRab is, in fact, an BL Her-type star not belonging to the cluster.   Making use of the $Gaia$-DR2 proper motions, we identified 5012 stars in the field of the cluster, that are very likely cluster members, and for which we possess photometry, enabling the production of a refined Colour-Magnitude Diagram. This also allowed us to identify a few variable stars that do not belong to the cluster. The RR Lyrae pulsation modes on the HB are cleanly separated by the first overtone red edge, a common feature in all Oo II type clusters. 
\end{abstract}


%%%%%%%%%%%%%%%%%%%%%%%%%%%%%%%%%%%%%%%%%%%%%%%%%%

%%%%%%%%%%%%%%%%% BODY OF PAPER %%%%%%%%%%%%%%%%%%

\section{Introduction}
M92 (NGC 6341) is one of the brightest globular clusters in the northern hemisphere; it is located in the constellation of Hercules ($\alpha = 17^{h}17'07.39'', \delta = +43^{\circ} 08' 09.4''$, J2000). It was discovered by Johann E. Bode in 1777 and independently rediscovered by Charles Messier in 1781. It is among the most metal-poor globular clusters in our galaxy ([Fe/H]$\sim$ -2.3), and given its Galactic position ($l$ =  $68.34^{\circ}$ , $b$ = $34.86^{\circ}$) and distance to the Sun of about 8.3 kpc, its reddening is naturally small; $E(B-V)=0.02$ mag  \citep{Harris1996}. According to the cluster Galactic orbit calculated by \citet{Baumgardt2019} with the \citet{Irrgang2013} Galactic model, M92 is approaching its apogalacticon, at about 10 kpc;  presently its Galactocentric distance is 8.4 $\pm$ 0.3 kpc. The mean proper motion components, calculated from the \citet{Gaia2018} ($Gaia$-DR2) are $\mu_\alpha cos (\delta)= -4.93 \pm 0.02$ mas/yr and $\mu_\delta =-0.57 \pm 0.02$ mas/yr \citep{Baumgardt2019}.

Time-series photometry of variable star populations in globular clusters has historically been of substantial relevance in the understanding of cluster properties, since the RR Lyrae stars were early recognised as standard candles for distance determinations  \citep[e.g.][]{Shapley1918}. Much later in the twentieth century, the SX Phe were also identified as distance indicators through their P-L relation \citep[e.g.][]{Nemec1993} and the RR Lyrae as metallicity indicators through their light curve morphology  and empirical calibrations \citep[e.g.][]{Kovacs1998}.
Hence, keeping our knowledge of the variable star population in the Galactic globular cluster system tidy, complete and properly classified, helps in achieving the fundamental goal of observational astronomy, which is transforming observational quantities into physical parameters. It is in this scope that the present paper is framed.

The known variable star population of M92, prior to the present investigation,
as listed in the Catalogue of Variable Stars in Globular Clusters (CVSGC: 2018 edition) \citep{Clement2001}, is composed by 17 RR Lyrae, 7 SX Phe stars, one anomalous cepheid (AC) (V7) and one eclipsing binary of the W UMa type (V14).  Thirteen of the 39 stars with a variable star designation and listed in the CVSGC, are not variable.
A CCD study of the variable star light curves in M92, their periods and pulsation mode composition in RR Lyrae and SX Phe stars was published by \citet{Kopacki2001} and \citet{Kopacki2007}, where a recount of the history of variable star discoveries in M92 is offered.
In spite of M92 being one of the brightest clusters in the northern hemisphere, there is not a detailed analysis of the light curves of its variable stars aimed to determine its physical parameters. Also, a reconsideration of its variable stars, based on a rather extensive time-series, in some cases extending the time-base to eighteen years, should enable refinement of periods and revision of classifications. The availability of state-of-the-art proper motions from the $Gaia$-DR2 opened the possibility to discuss
the stellar cluster membership with  unprecedented detail, not only for stars measured in the field of a given cluster, but specially for the detected variable stars.

The present paper is part of a series of studies of the variable star populations in selected globular clusters, based on \emph{VI} CCD time series.
We aim to calculate individual physical parameters (luminosity, mass, effective temperature, radii and metallicity), from the light curve morphology of RR Lyrae stars via the Fourier decomposition. This approach and the P-L relations of RR Lyrae and SX Phe stars yield the mean distance of the parental cluster. We shall explore the field of the cluster for non-members and use a clean version of the Colour-Magnitude Diagram (CMD) to discuss the distributions of RR Lyrae stars on the Horizontal Branch (HB) and the SX Phe stars among the blue stragglers region. 

The paper is organized in the following way: in $\S$ 2 the observations and the transformation to the standard system are described; in $\S$ 3 the stellar membership analysis and the approach to the cluster membership of variable stars is presented; in $\S$ 4 the refined CMD of M92, is presented along with the variable stars, isochrones and Zero-Age Horizontal Branch (ZAHB); in $\S$ 5 the period calculation and a search for new variables is described and a discussion is given on the multi-mode detection in some variables; in $\S$ 6 the Fourier decomposition towards the estimation of physical parameters is discussed; in $\S$ 7 we summarise our results and conclusions. Appendix A contains a brief discussion of peculiarities on individual variables.

\section{Observations}

The observations were performed during 11 nights between June 10th and August 15th, 2018 with the 0.84 m. telescope at the Observatorio Astron\'omico Nacional on the Sierra San Pedro M\'artir (OAN-SPM), Baja California, M\'exico. A total of 649 and 753 images were obtained  in the Johnson-Kron-Cousins (JKC) $V$ and $I$ filters, respectively. The detector was a Spectral Instruments CCD of 1024$\times$1024 pixels with a scale of 0.444 arcsec/pix, which translates into a field of view (FoV) of approximately 7.57$\times$7.57~arcmin$^2$. The log of observations is given in Table \ref{tab:observations} where the dates, number of frames, exposure times and average nightly seeing are recorded. 

\begin{table}
\footnotesize
\caption{Observations log of M92. Data are from San Pedro M\'artir (SPM). Columns $N_{V}$ and $N_{I}$ correspond to the number of images taken with the $V$ and $I$ filters respectively. Columns $t_{V}$ and $t_{I}$ provide the exposure times. In the last column the average seeing is listed.}
\centering
\begin{tabular}{lccccc}
\hline
Date & $N_{V}$ & $t_{V}$ (s) & $N_{I}$ &$t_{I}$ (s)&seeing (") \\
\hline
 20180610 & 168 & 50 & 184 & 30 & 2.1 \\
 20180611 & 120 & 50 & 150 & 30 & 1.7 \\
 20180613 &  36 & 60 &  41 & 40 & 1.9 \\
 20180716 &  48 & 60 &  56 & 40 & 1.3 \\
 20180724 &  17 & 60 &  21 & 40 & 1.5 \\
 20180425 &  50 & 60 &  53 & 40 & 1.6 \\
 20180728 &  58 & 60 &  60 & 40 & 1.8 \\
 20180730 &  26 & 60 &  32 & 40 & 1.6 \\
 20180731 &  32 & 60 &  35 & 40 & 1.3 \\
 20180801 &  41 & 60 &  50 & 40 & 1.5 \\
 20180815 &  53 & 60 &  71 & 40 & 1.8 \\
\hline
Total:    & 649 & -- & 753 & -- & --\\
\hline
\end{tabular}
\label{tab:observations}
\end{table}

M92 was also observed between August and September, 2000 and May, 2001 with the 1.5 m. telescope at the OAN-SPM. The images were obtain in the Johnson $V$ filter. The data were used in a Master thesis \citep{Marin2002} and were never published. In the present paper we shall use the data of those variables included in that study, all RR Lyrae stars, as they complement very well our present data and extend the time-base to eighteen years. 

\subsection{Difference image analysis}
We used the Difference Image Analysis (DIA) technique with its pipeline DanDIA \citep{Bramich2008,Bramich2013} to obtain high-precision photometry for all the point sources in the FoV of our images of M92. The strategy consists on creating a reference image by stacking the best quality images in each filter and then sequences of differential images in each filter were built by subtracting the respective reference image from the rest of the collection. Differential fluxes for each star detected in the reference image were then measured on each differential image. Light curves for each star were constructed by calculating the total flux $f_{\mbox{\scriptsize tot}}(t)$ in ADU/s at each epoch $t$ from:
\begin{equation}
f_{\mbox{\scriptsize tot}}(t) = f_{\mbox{\scriptsize ref}} + \frac{f_{\mbox{\scriptsize diff}}(t)}{p(t)},
\label{eqn:totflux}
\end{equation}
where $f_{\mbox{\scriptsize ref}}$ is the reference flux (ADU/s), $f_{\mbox{\scriptsize diff}}(t)$ is the differential flux (ADU/s) and $p(t)$ is the photometric scale factor (the integral of the kernel solution). Conversion to instrumental magnitudes was achieved using:
\begin{equation}
m_{\mbox{\scriptsize ins}}(t) = 25.0 - 2.5 \log \left[ f_{\mbox{\scriptsize tot}}(t)
\right],
\label{eqn:mag}
\end{equation}
where $m_{\mbox{\scriptsize ins}}(t)$ is the instrumental magnitude of the star at time $t$. The above procedure has been described in detail in \citet{Bramich2011}.

\begin{figure} 
\includegraphics[width=8.0cm,height=8.0cm]{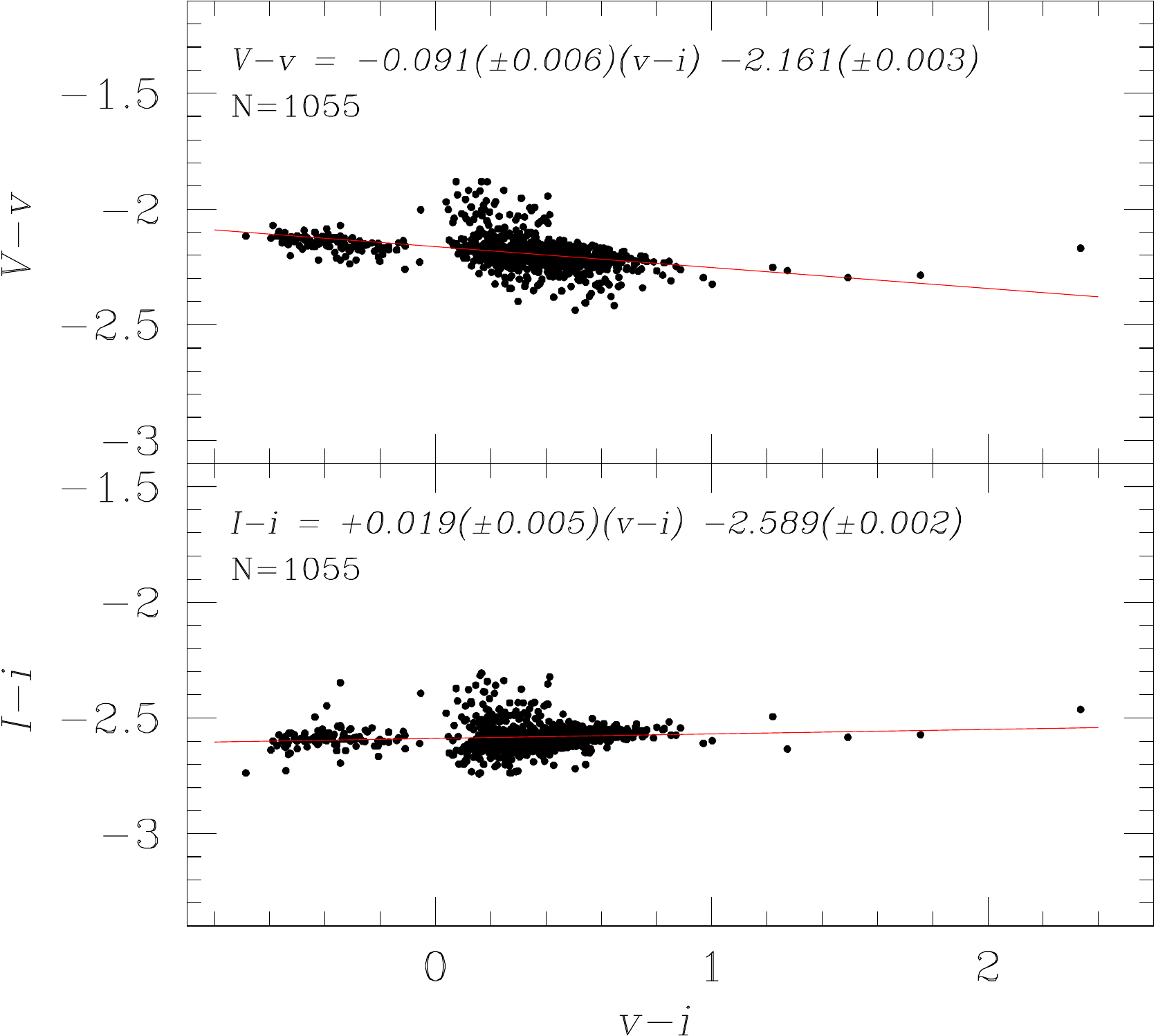}
\caption{The instrumental-to-standard photometric systems transformation equations for the  $V$ and $I$ filters. These equations were calculated based on 1055 local standard stars from the collection of \citet{Stetson2000} in the FoV of our images.}  
    \label{trans}
\end{figure}

\subsection{Transformation to the standard system}
\label{calibration}

We used the standard stars in the FoV of M92 included in the online collection of \citet{Stetson2000} to transform instrumental $vi$ magnitudes into the JKC standard $VI$ system. We identified a total of 1055 stars with $V$ and $I$ magnitudes. The mild colour dependence of the standard minus instrumental magnitudes is shown in Fig. \ref{trans}, the transformation equations between the instrumental and the standard magnitudes are given in each panel of the figure. We may note in Fig. \ref{trans} a group of rather discrepant stars above the main trends, within  $0.<(v-i)<0.4$. These are mostly stars near the turn-off point at the faint end of our photometric capabilities, hence  with lower precision. Although they have been included in the fit, they represent less about 4\% of the sample and removing them does not have a significant impact on the transformation to the standard system.

\begin{figure}
\begin{center}
\includegraphics[width=8cm]{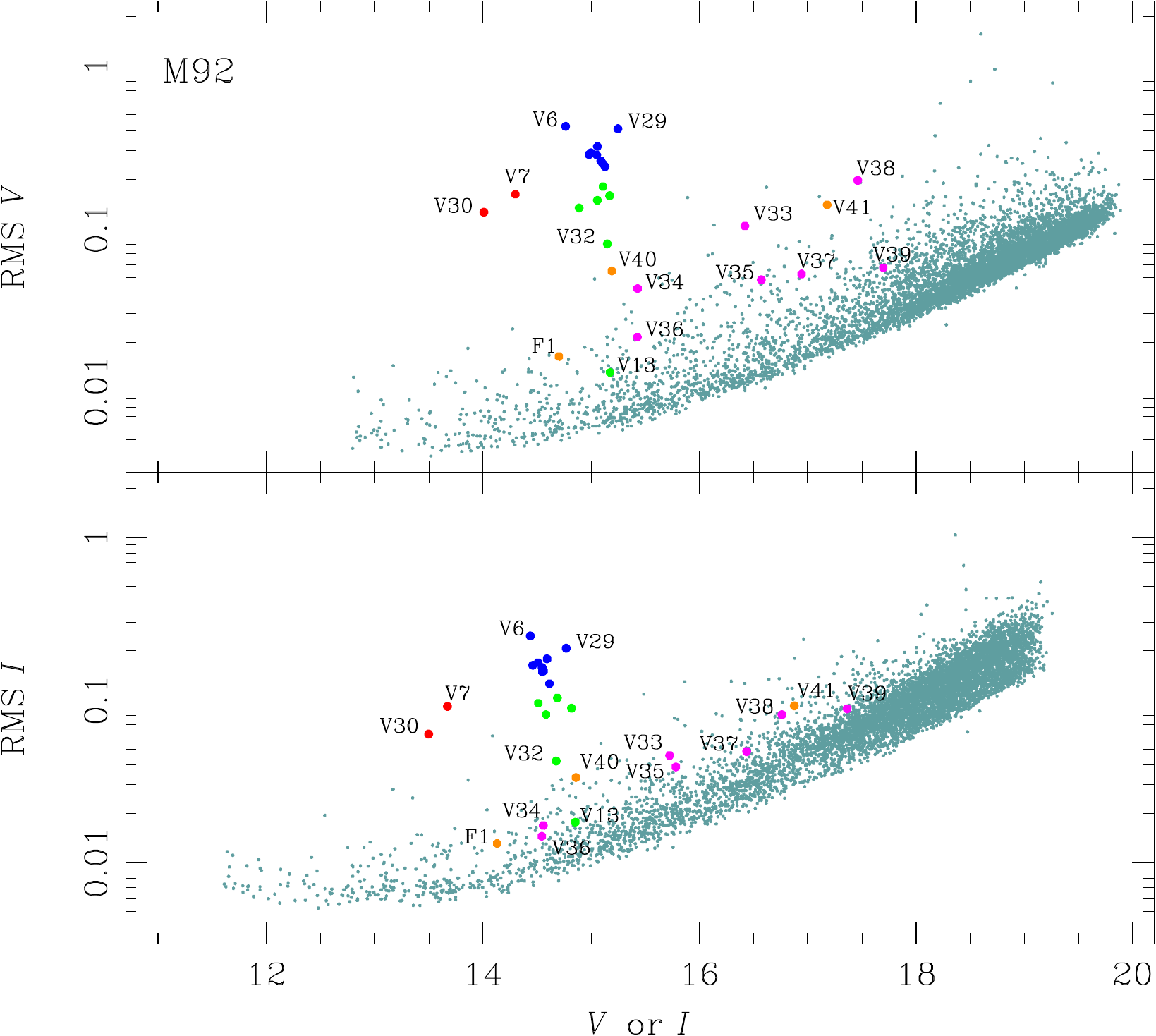}
\caption{The rms magnitude deviations as a function of the mean magnitudes $V$ (upper panel) and $I$ (lower panel).  The blue and green circles correspond to RRab and RRc stars respectively, red circles identify
stars above the HB. Pink circles identify SX Phe stars and yellow symbols mark the newly identified variables in this work. Cyan dots are non-variable stars measured in the FoV.}
\label{rms_M92}
\end{center}
\end{figure}

The precision of our photometry can be judged from the rms vs. magnitude diagrams in Fig. \ref{rms_M92} where the variable stars, naturally outstanding with larger rms values, are indicated.

The time-series \emph{VI} photometry obtained in this work for all variable stars is reported in Table \ref{tab:vi_phot}. For completeness we include the unpublished data from the MSc thesis of Zaida Mar\'in \citep{Marin2002}. Only a small portion of this table is included in the printed version of the paper. The full table shall be available in electronic form in the Centre de Donnés astronomiques de Strasbourg database (CDS).

\begin{table}
\scriptsize
\begin{center}
\caption{Time-series \textit{VI} photometry for the variable stars observed in this work$^*$. In the electronic version, data from \protect\citet{Marin2002} are included.}
\label{tab:vi_phot}
\centering
\begin{tabular}{cccccc}
\hline
Variable &Filter & HJD & $M_{\mbox{\scriptsize std}}$ &
$m_{\mbox{\scriptsize ins}}$
& $\sigma_{m}$ \\
Star ID  &    & (d) & (mag)     & (mag)   & (mag) \\
\hline
 V1 & $V$& 2458279.66953& 15.052 & 17.224 & 0.005 \\   
 V1 & $V$& 2458279.67030& 15.039 & 17.212 & 0.005 \\
\vdots   &  \vdots  & \vdots & \vdots & \vdots & \vdots  \\
 V1 & $I$ & 2458279.66677 & 14.469 & 17.055 &  0.004\\  
 V1 & $I$ & 2458279.66730 & 14.481 & 17.067 & 0.004 \\ 
\vdots   &  \vdots  & \vdots & \vdots & \vdots & \vdots  \\
 V2 & $V$ & 2458279.66953 & 14.905& 17.074 & 0.005 \\   
 V2 & $V$ & 2458279.670301 & 14.913& 17.082 &  0.005 \\
\vdots   &  \vdots  & \vdots & \vdots & \vdots & \vdots  \\
 V2 & $I$ & 2458279.66677 & 14.440&  17.027 & 0.006 \\    
 V2 & $I$ & 2458279.66732 & 14.438&  17.025 & 0.006 \\   
\vdots   &  \vdots  & \vdots & \vdots & \vdots & \vdots  \\
\hline
\end{tabular}
\end{center}
* The standard and instrumental magnitudes are listed in columns 4 and~5, respectively. Filter and epoch of mid-exposure are listed in columns 2 and 3, respectively. The uncertainty on $\mathrm{m}_\mathrm{ins}$ is listed in column~6, which also corresponds to the uncertainty on $\mathrm{M}_\mathrm{std}$. A full version of this table is available on the CDS database. For V42 and C1 \emph{vi} are instrumental magnitudes.

\end{table}

\section{Stellar membership in the field of M92}
\label{members}

We have employed proper motions of the $Gaia$-DR2 and the method of \citet{Bustos_Fierro2019}. This method consists of two stages: the first stage is based on the Balanced Iterative Reducing and Clustering using Hierarchies (BIRCH) algorithm \citep{Zhang1996} in a four-dimensional space of physical parameters-gnomonic projection of celestial coordinates and proper motions- that detects groups of stars in this 4D space; in the second stage an analysis of the projected spatial distribution of stars with different proper motions allows the extraction of most of the members in the outskirts of the cluster or with large proper motions dispersion. Out of the 6974 stars measured by our photometric approach, 5012 were identified as very likely members following this method. This enabled us to refine the CMD and to discuss the cluster membership of specific variable stars.

\subsection{On the cluster membership of know variable stars.}

\begin{figure*}
\centerline{\includegraphics[width=18cm]{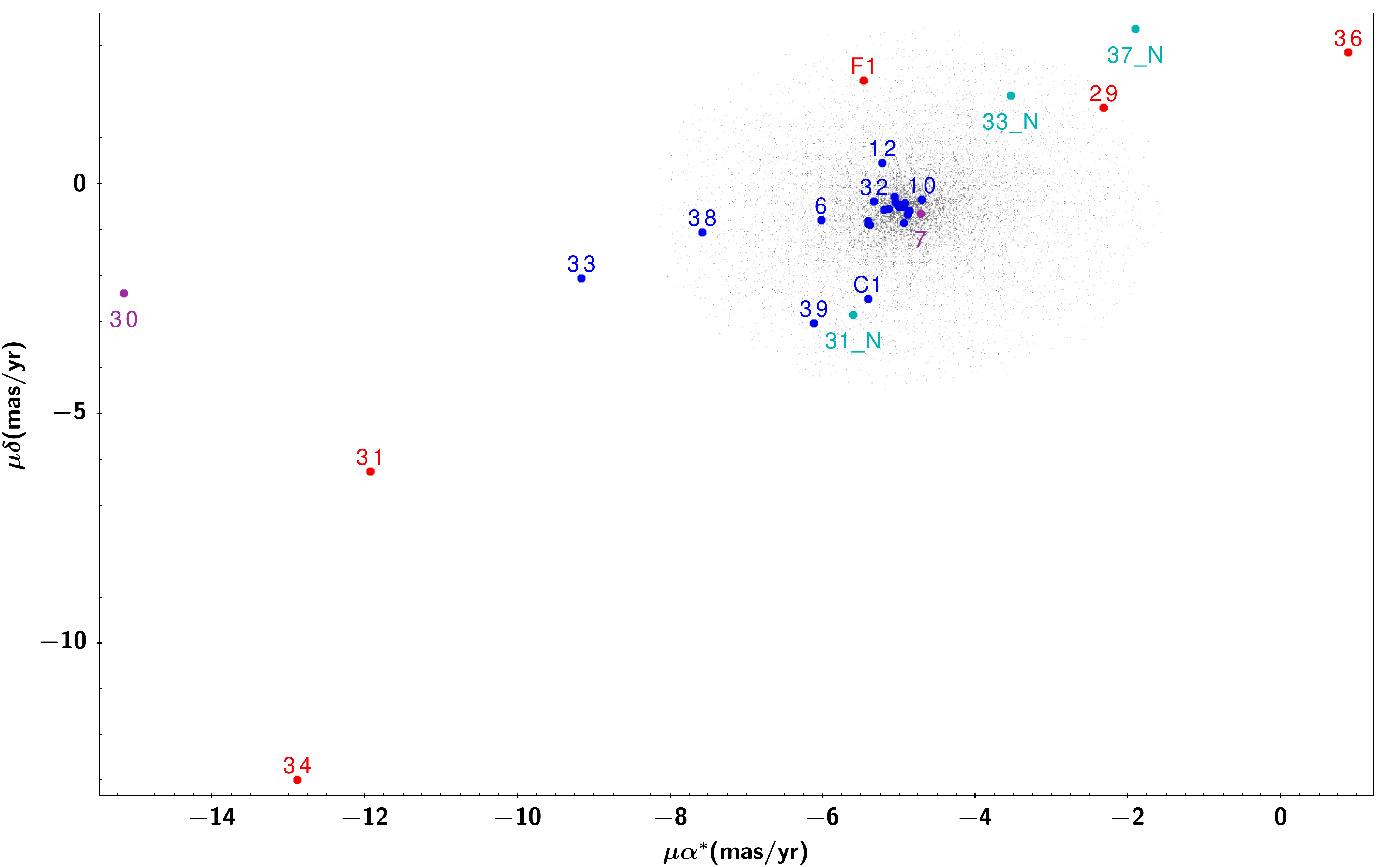}}
\caption{VPD of the stars in the central region of M92. Light grey dots are stars considered cluster members (see $\S$ \ref{members}).
Coloured symbols are used for variable stars; blue for stars that are considered to belong to the cluster, red for stars with clearly discrepant proper motions; these stars are likely non-cluster members and their position in the CMD is equally peculiar, as it is discussed in Appendix A. A few stars have two $Gaia$ sources within the PSF of our photometry, the proper motion of the secondary star is marked with light-blue colour. The AC star (V7) and BL Her-type star (V30) are indicated with purple colour.}
\label{PMM92}
\end{figure*}

In Fig. \ref{PMM92} the proper motions of all known variables in the FoV of our images of M92, taken from $Gaia$-DR2, are displayed. In this figure, the proper motion of variable stars are compared with the member star population. It is clear that a few variables do not participate of the general cluster motion, indicating that they are likely not cluster members. These and their peculiar position in the CMD (inconsistent with their variable type), reinforces the idea that they do not belong to the cluster. Individual cases shall be discussed in Appendix  \ref{IndStars}.

\section{The Colour-Magnitude Diagram of M92}
\label{sec:CMD}
\begin{figure*}
\begin{center}
\includegraphics[width=15cm]{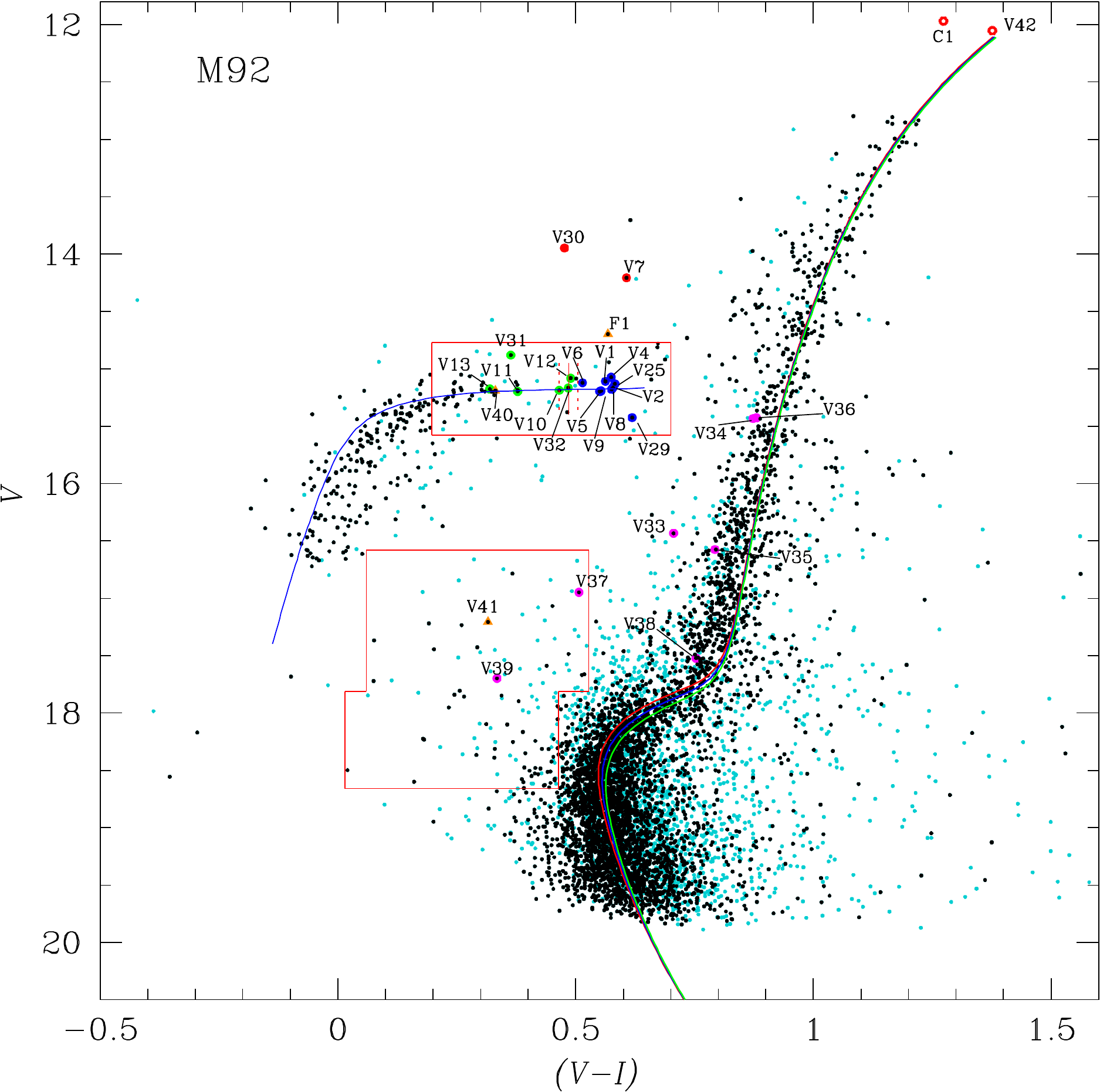}
\caption{CMD of M92. The colors are: blue for the RRab stars, green for the RRc stars, pink for the SX Phe and the red circle corresponds to the stars well above the HB (V7, V30). Orange triangles are variables newly discovered in this work. Empty red circles for V42 and C1 were placed adopting the  $Gaia$-DR2 photometric indices and transforming them into the JKC system. The black dots represent cluster members whereas cyan dots are used for field stars. The isochrones for ages of 12.0 Gyr (red), 12.5 Gyrs (blue) and 13.0 Gyrs (green) are from \citet{VandenBerg2014} (see $\S$\ref{sec:CMD} for details). The continuous red vertical line in the HB, represents the red edge of the first overtone instability strip duly reddened, with an uncertainty of $\pm 0.02$ mag indicated by the dashed vertical segments \citep{Arellano2016}. The red boxes define the regions where a search of new variables was performed.}
\label{CMD_M92}
\end{center}
\end{figure*}

The CMD of M92 is shown in the Fig. \ref{CMD_M92}, the location of 29 variables star is marked, the color code is explained in the figure caption. All variable stars are plotted using their intensity-weighted means $<V>$ and corresponding colour $<V> - <I>$. The black dots are cluster member stars selected by the method of \citet{Bustos_Fierro2019} using the proper motions at the $Gaia$-DR2 database. The black vertical line in the HB, represents the border between the fundamental and first overtone pulsating modes in RR Lyrae stars, i.e. the red edge of the first overtone (FORE) instability strip \citep{Arellano2016}, duly reddened for M92. It is clear that in M92 the fundamental and first overtone pulsators, RRab and RRc respectively, are well segregated by the FORE, a characteristic of all Oo II type clusters. This will be further addressed below in $\S$ \ref{conclusion}.

In the plot, we include three isochrones of ages 12.0 Gyrs (red), 12.5 Gyrs (blue) and 13.0 Gyrs (green), and ZAHB of [Fe/H]=-2.2, Y=0.25 and [$\alpha$/Fe]=0.4; all these models come from the Victoria-Regina model collection of \citet{VandenBerg2014}. Isochrones and ZAHB were shifted to a distance of 8.47 kpc calculated for the RRab stars and a reddening of $E(B-V)=0.02$.

\section{Variable stars}
\begin{figure*} 
\includegraphics[width=17.1cm,height=18.5cm]{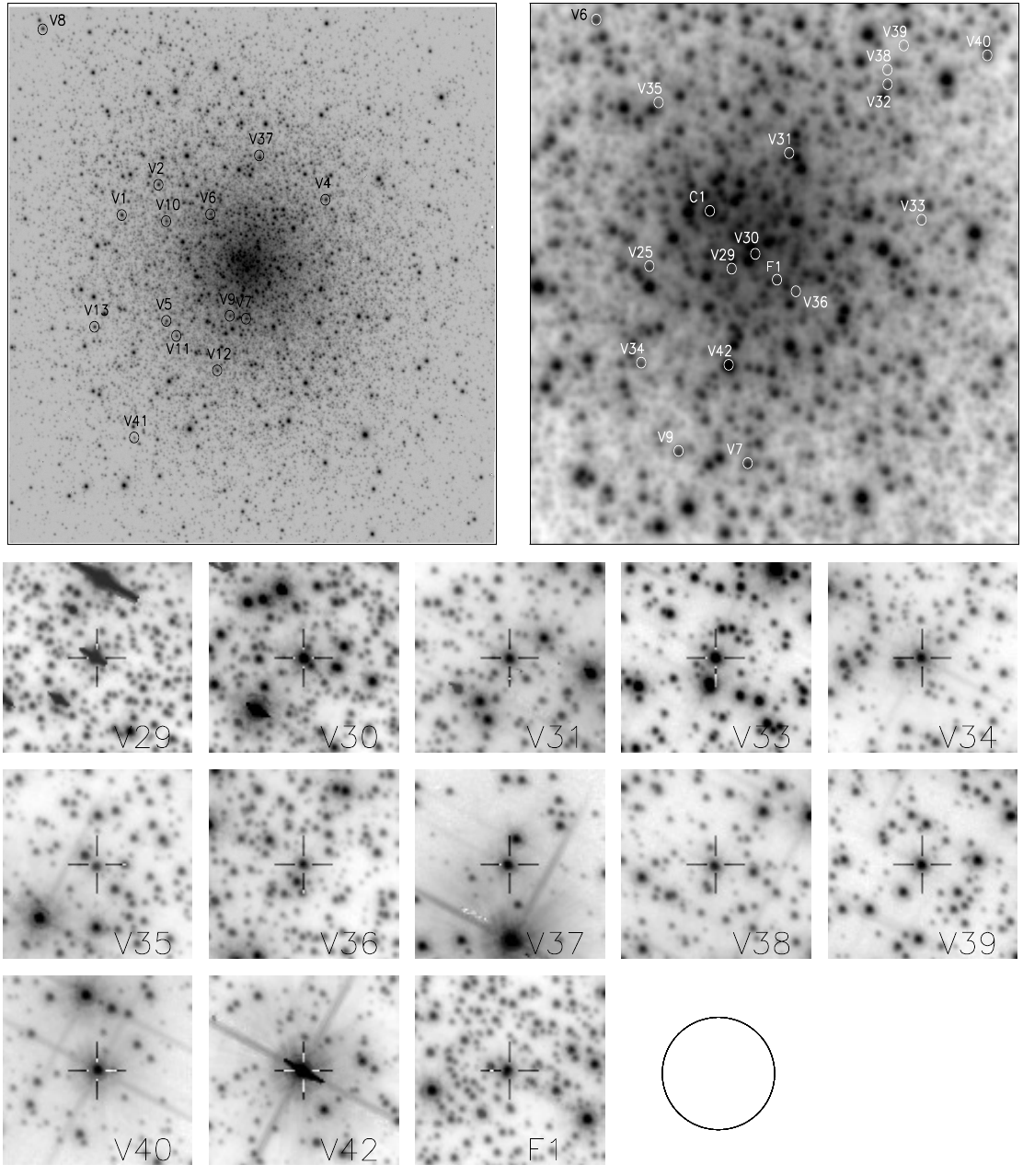}
\caption{Identification charts of all known variables in M92. The upper two panels are based on the $V$ reference image built by stacking several of the best quality images in our collection. The panel to the left is a $7.57\times7.57$ arcmin$^2$ field while the panel to the right includes the central $1.78\times1.78$ arcmin$^2$. For better identification, particularly of faint stars in crowded regions, the lower panels are cuts of a high resolution Hubble Space Telescope image. The size of these smaller cuts is about $5.0\times5.0$ arcsec$^2$. V30, is a BL Her star that is not a cluster member. V33 and V37 are SX Phe stars but evidence is found in favour of V33 not belonging to the cluster (see Appendix \ref{IndStars}).
 In all the images the North is up and East is to the left. The black circle is approximately the size of our photometric psf, which makes clear that in some cases we have measured two (or more) unresolved stars in our images.} 
\label{ID_CHART}
\end{figure*}

 All known variable stars in the FoV of the 2018 observations, are identified in the charts of Fig. \ref{ID_CHART}. The stars V3 and V14 are off those limits. However, available data for V3 from \citep{Marin2002} enabled us to include the star in our discussion.

\begin{table*}
%\scriptsize
\begin{center}
\caption{Data of Variable stars in M92 in the FoV of our images.}
\label{tab:variables}
\begin{tabular}{llcccclccccc}
\hline
Variable & Variable & $<V>$ & $<I>$ & $A_V$ & $A_I$ & P (days) & P (days) & HJD$_{\rm max}$ &  RA  &  Dec. & $Gaia$ Source ID\\
Star ID  &  Type  & (mag) & (mag) & (mag) & (mag) & this work & CVSGC & (2450000+)  & (J2000.0) & (J2000.0) & (1.36040$\times10^{18}$+)\\
\hline
V1  & RRab           & 15.109 & 14.547 & 0.905 & 0.556 &  0.702786 & 0.7028 & 8280.8720 & 17:17:18.96 & +43:08:54.68 & 5567883886720 \\
V2  & RRab \emph{Bl} & 15.163 & 14.584 & 0.935 & 0.534 &  0.643888 & 0.6439 & 8280.8451 & 17:17:15.62 & +43:09:22.06 & 5602243616640 \\
V3  & RRab           & 15.147 &    --   & 1.130 &   --  &  0.637497 & 0.6377 &     --    & 17:17:12.00 & +43:12:25.30  & 8905076366208 \\
V4  & RRab           & 15.073 & 14.499 & 0.997 & 0.573 &  0.628934 & 0.6289 & 8280.8712 & 17:17:00.34 & +43:09:08.72 & 5366019655040 \\
V5  & RRab           & 15.193 & 14.640 & 0.999 & 0.617 &  0.619685 & 0.6197 & 8282.8301 & 17:17:14.88 & +43:07:19.12 & 4781907957504 \\
V6  & RRab           & 15.122 & 14.608 & 1.146 & 0.718 &  0.600038 & 0.6000 & 8280.9510 & 17:17:10.86 & +43:08:55.53 & 5499164380032 \\
V7  & AC             & 14.207 & 13.600 & 0.679 & 0.420 &  1.061150 & 1.0614 & 8324.8088 & 17:17:07.59 & +43:07:21.18 & 5396085147008 \\
V8  & RRab \emph{Bl} & 15.179 & 14.603 & 0.965 & 0.587 &  0.673172 & 0.6728 & 8345.8322 & 17:17:26.23 & +43:11:42.32 & 7324528224512 \\
V9  & RRab \emph{Bl} & 15.198 & 14.647 & 1.118 & 0.659 &  0.609370 & 0.6085 & 8324.8223 & 17:17:09.08 & +43:07:23.77 & 5430444893568 \\
V10 & RRc            & 15.187 & 14.722 & 0.516 & 0.296 &  0.377279 & 0.3773 & 8279.7663 & 17:17:14.91 & +43:08:49.17 & 5533524137216 \\
V11 & RRc \emph{Bl} $^{1}$     & 15.197 & 14.819 & 0.479 & 0.277 &  0.308454 & 0.3084 & 8330.7141 & 17:17:13.96 & +43:07:05.54 & 4610108446464 \\
    &                &        &        &       &       &  0.182553 &        &           &             &              &               \\
V12 & RRc            & 15.081 & 14.592 & 0.429 & 0.245 &  0.409916 & 0.4100 & 8280.8437 & 17:17:10.22 & +43:06:34.39 & 4674530655744 \\
V13 & RRc            & 15.176 & 14.856 & 0.049 & 0.053 &  0.301536 & 0.3130 & 8345.8284 & 17:17:21.45 & +43:07:13.63 & 4541388893952 \\
    &                &        &        &       &       &  0.337182 &        &           &             &              &               \\
V25 & RRab           & 15.131 & 14.548 & 0.791 & 0.485 &  0.700404 & 0.7004 & 8280.7670 & 17:17:09.71 & +43:08:03.09 & 5434740074496 \\
V29$^{2}$ & RRab     & 15.425 & 14.806 & 1.233 & 0.744 &  0.595967 & 0.5959 & 8280.7850 & 17:17:07.94 & +43:08:02.56 & 5469102206720 \\
V30$^{2}$ & BL Her            & 13.947 & 13.471 & 0.521 & 0.237 &  0.528331 & 0.5283 & 8330.7804 & 17:17:07.48 & +43:08:06.00 & 5469102036480 \\
V31$^{2}$ & RRc      & 14.880 & 14.517 & 0.419 & 0.300 &  0.398340 & 0.3982 & 8279.8654 & 17:17:06.76 & +43:08:27.10 & 5469102216576 \\
V32 & RRc            & 15.166 & 14.682 & 0.227 & 0.122 &  0.324654 & 0.3246 & 8345.7216 & 17:17:04.59 & +43:08:41.83 & 5675260454656 \\
V33 & SX Phe         & 16.434 & 15.728 & 0.423 & 0.196 &  0.075094 & 0.0751 & 8329.8158 & 17:17:03.86 & +43:08:13.70 & 5675256740224 \\
V34$^{2}$ & SX Phe   & 15.433 & 14.560 & 0.170 & 0.078 &  0.083009 & 0.0830 & 8324.8049 & 17:17:09.89 & +43:07:42.70 & 5434742361344 \\
V35 & SX Phe         & 16.575 & 15.782 & 0.186 & 0.118 &  0.054909 & 0.0549 & 8330.8018 & 17:17:09.52 & +43:08:37.92 &               \\
V36$^{2}$ & SX Phe   & 15.426 & 14.546 & 0.088 & 0.055 &  0.047461 & 0.0475 & 8330.7784 & 17:17:06.55 & +43:07:58.10 & 5469102024064 \\
V37 & SX Phe         & 16.948 & 16.441 & 0.197 & 0.149 &  0.040194 & 0.0402 & 8327.7860 & 17:17:06.42 & +43:09:48.70 & 5743976585472 \\
V38 & SX Phe         & 17.522 & 16.769 & 0.732 & 0.360 &  0.072945 & 0.0729 & 8280.7930 & 17:17:04.59 & +43:08:44.91 & 5675260421888 \\
V39 & SX Phe         & 17.698 & 17.363 & 0.125 & 0.084 &  0.036980 & 0.0359 &    --     & 17:17:04.24 & +43:08:50.08 & 5675260460160 \\
V40 & RRd            & 15.191 & 14.860 & 0.174 & 0.128 &  0.276639 &        & 8280.7402 & 17:17:02.44 & +43:08:48.01 & 5743976516352 \\
    &                &        &        &       &       &  0.201314 &        &           &             &              &               \\
V41 & SX Phe         & 17.207 & 16.891 & 0.650 & 0.503 &  0.055947 &        & 8315.7914 & 17:17:17.80 & +43:05:33.71 & 4434010244864 \\
    &                &        &        &       &       &  0.054350 &        &           &             &              &               \\
    &                &        &        &       &       &  0.057632 &        &           &             &              &               \\
V42 & SR ?           & 12.052$^{3}$ & 10.676$^{3}$ & 0.180 &   --  &   14.38   &        &    --     & 17:17:08.00 & +43:07:42.10 & 5400382533376 \\
F1$^{2}$ & RRc      & 14.697 & 14.130 & 0.069 & 0.051 &  0.334540 &        & 8329.8702 & 17:17:06.97 & +43:08:00.22 & 5469102053504 \\
C1  & SR ?           & 11.969$^{3}$ & 10.696$^{3}$ & 0.100 &   --  &   21.10   &        &    --     & 17:17:08.41 & +43:08:14.84 & 5503461790848 \\

\hline
\end{tabular}
\raggedright
\center{\quad \emph{Bl}: RR Lyrae with Blazhko effect.
1. The double mode of this star is not confirmed.
2. Not a cluster member.
3. Magnitude obtained from the transformations between $Gaia$-DR2 and the JKC system}
\end{center}
\end{table*}

\begin{figure}
\centerline{\includegraphics[width=8cm]{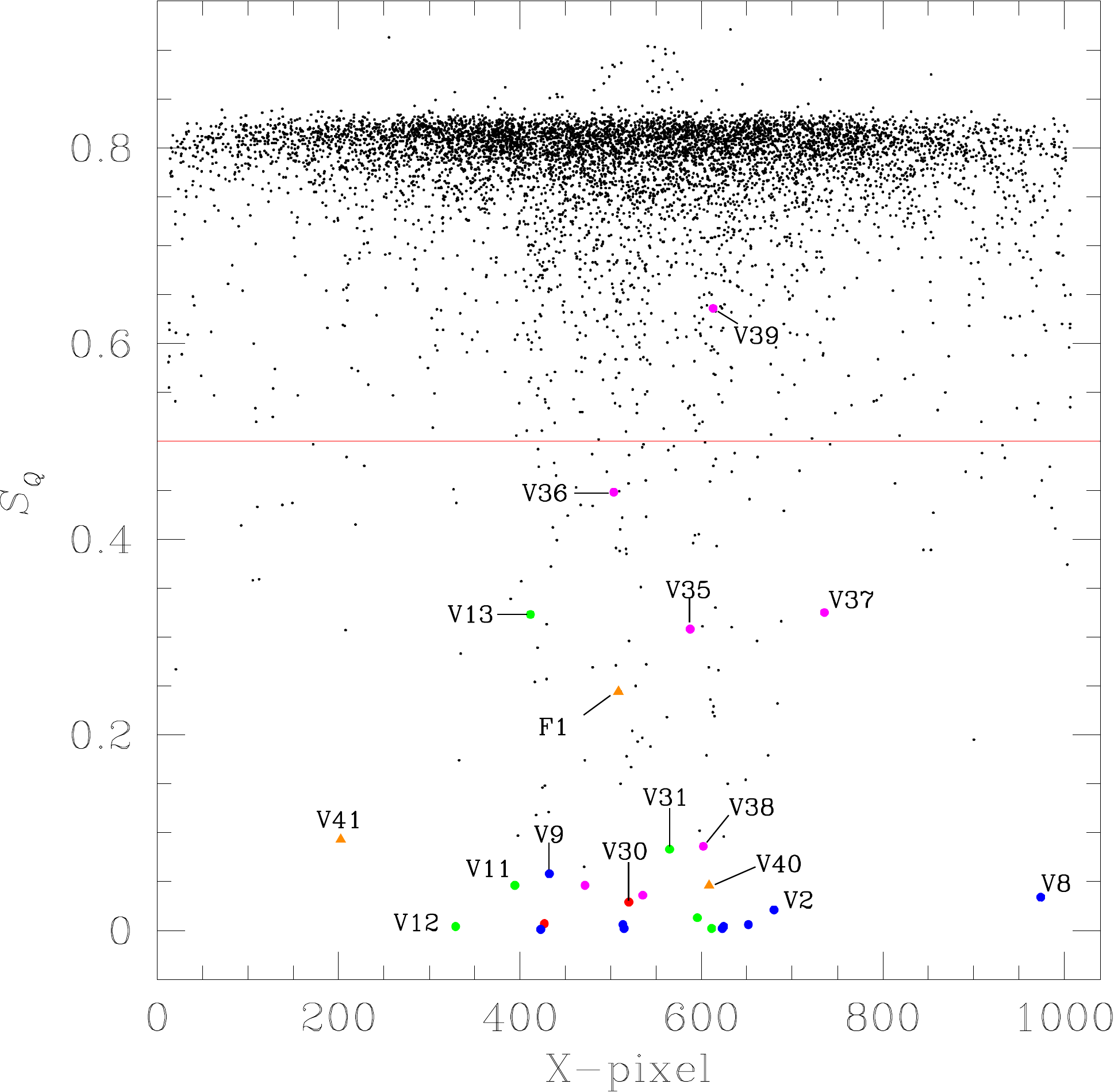}}
\caption{Plane $S_Q$ versus the X-coordinate on the CCD. A total of 6973 stars are plotted. We choose as an arbitrary threshold $S_Q$=0.5. We realised a search of variability for all star below this value.}
\label{SQ}
\end{figure}

\subsection{Period determination}
For seven variables, named V1, V2, V5, V8, V10, V11 and V12, we have data spanning 18 years, which enable to substantially refine their periods.
For the rest of the sample,  we only dispose of data from 2000-2001 (for V3) or 2018. The periods were estimated using the string-length method \citep{Burke1970,Dworetsky1983} and Period04 code \citep{Lenz2005}, and are listed in Table \ref{tab:variables}. For comparison we include in column 8 the periods reported in the CVSGC.

\begin{figure*}
\centerline{\includegraphics[width=16cm]{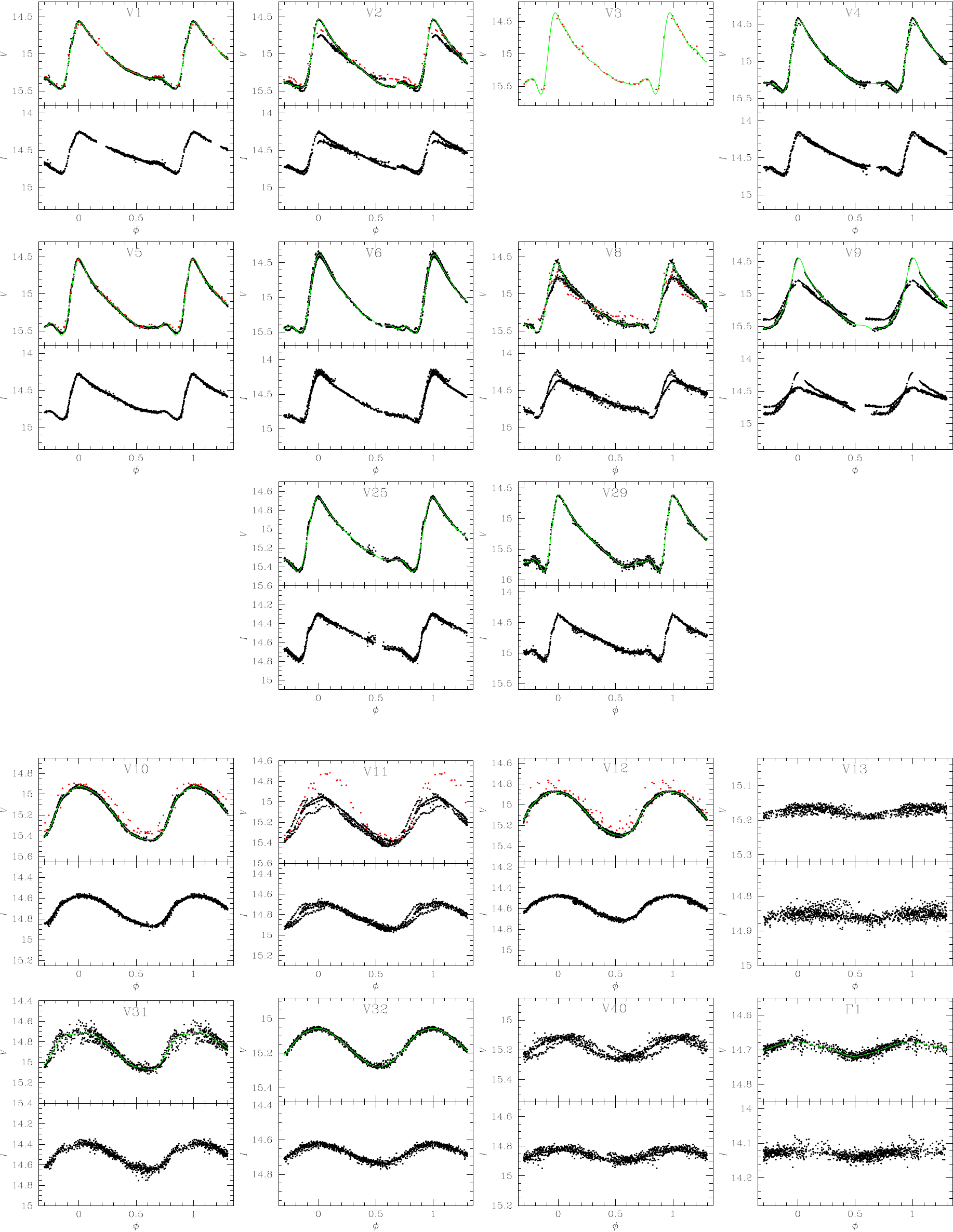}}
\caption{\emph{VI} light curves of RR Lyrae stars in the field of M92. Black symbols correspond to 2018 data, and red symbols are employed for 2000-2001 data from \citet{Marin2002}. Fourier fits, whose parameters are given in Table \ref{tab:fourier_coeffs} (see $\S$ \ref{FOURIER}), are shown in green colour. Note that for V3 we only have data from \citet{Marin2002}.}
\label{mosaico_RR}
\end{figure*}

\subsection{Search for new variables}

We have taken advantage of the time-series  images to search for new variables in our FoV. We performed the search by three different strategies. First we used the string-length method, in which each light curve was phased with periods between 0.02 and 1.7 d calculating a normalised statistic value $S_Q$. This value measures the minimum dispersion of the phased light curve. The plane $S_Q$ versus the X-coordinate for each star is shown in Fig \ref{SQ}. We have drawn an arbitrary threshold at $S_Q=0.5$, below which we found almost all the known variables with the exception of V39, as their dispersion for a correct period is minimum. The light curves of each star below this threshold were explored individually and found three new variables, labeled V40, V41 and F1; of types RRd, SX Phe and RRc respectivelly.

As a second approach, we isolated the light curves within the regions defined by the red contours in the CMD, since these are regions where variable star are commonly found (RR Lyrae in the HB and SX Phe in the Blue Stragglers region). Although no new variables were found, we corroborated the variability of the three stars found by the previous method.

Finally, we performed a search by blinking of all the residual images in our collection, and confirmed the variability of all previously known variables, the three new ones reported above and found a new variable and a probable variable, both lying near the tip of the red giant branch (RGB). They were labeled V42 and C1. Since they are saturated in the $I$-band, we only have instrumental light curves for these stars. We placed them in the CMD by adopting the $Gaia$-DR2 photometric indices and transforming them into the JKC system. These are near the tip of the RGB and are likely SR stars, but a longer time span would be required to confirm their variations and classification.

\begin{figure}
\centerline{\includegraphics[width=8.3cm]{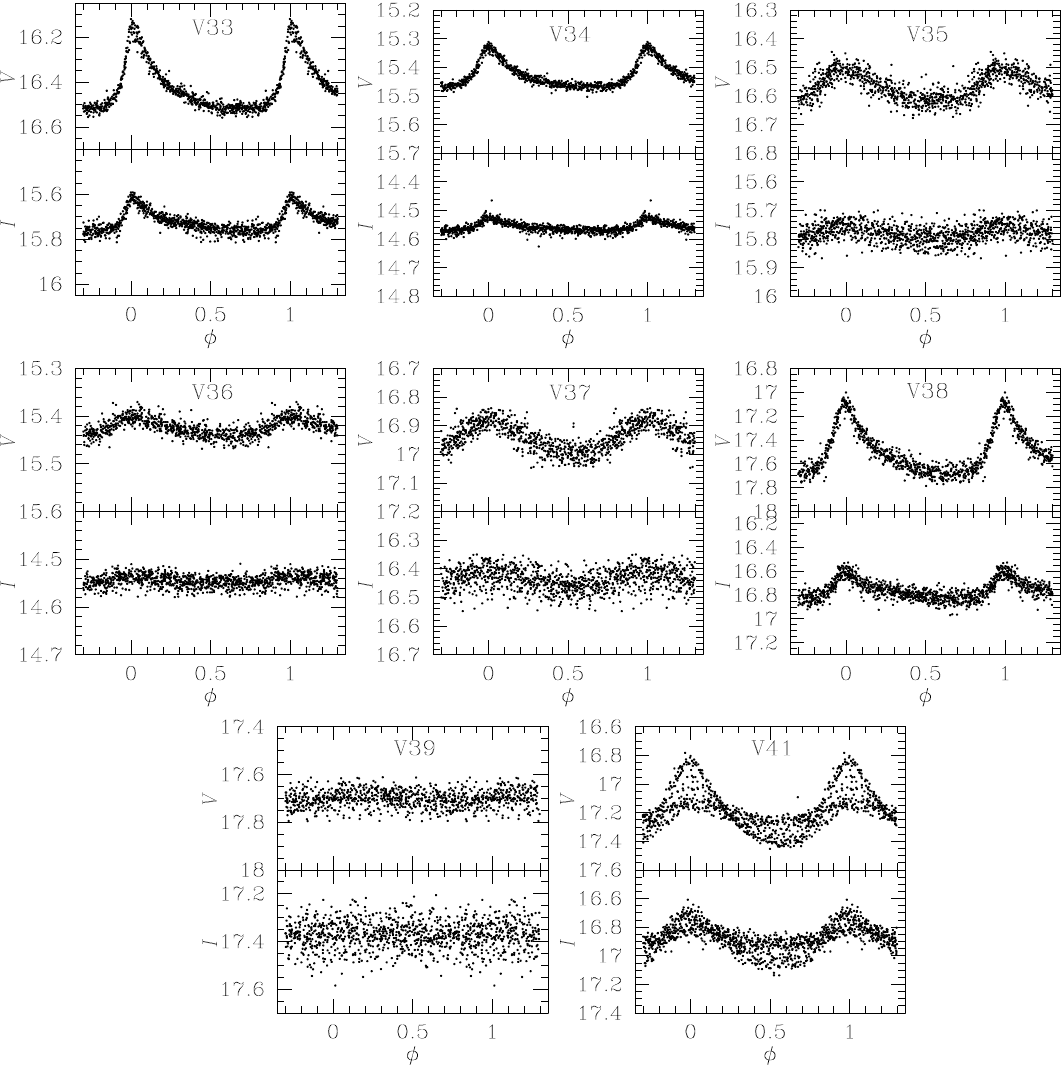}}
\caption{Light curves in $V$ and $I$ filters of SX Phe stars in M92. V41 is a new discovery in this work for which we found three active frequencies (see Appendix \ref{IndStars}.)}
\label{mosaicoSX}
\end{figure}

\begin{figure}
\centerline{\includegraphics[width=8.3cm]{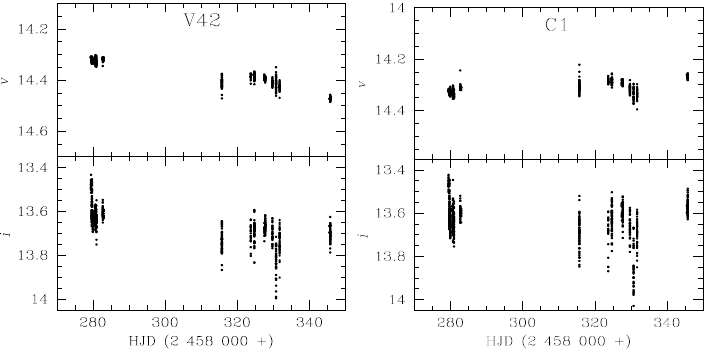}}
\caption{Instrumental \emph{vi} light curves of the stars V42 and C1. They both seem to belong to the cluster and sit near the tip of the RGB. They are very likely to be SR-type stars.}
\label{SR}
\end{figure}

\begin{figure}
\includegraphics[width=8.3cm]{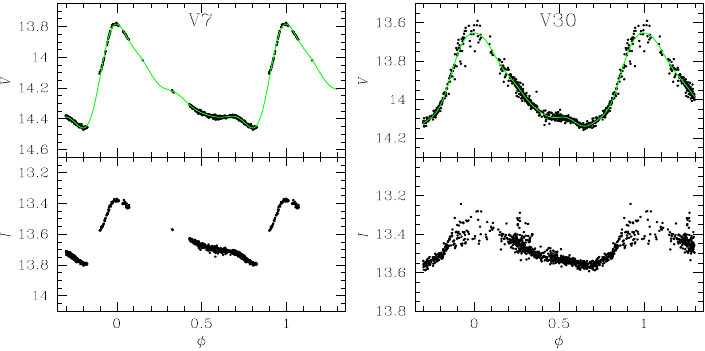}
\caption {Light curves of stars above the HB. V7 is phased with a period of 1.06115 d and V30 with 0.528331 d. V7 is confirmed as an AC star. V30 however, is classified as BL Her-type star. (see Appendix \ref{IndStars}.)}
\label{mosaico_AC}
\end{figure}

After this search, there are 30 confirmed variables in the field of M92 and one that should remain as a candidate variable until confirmed. Seven of these stars are not cluster members. The light curves of all variables in our FoV are displayed in Figs. \ref{mosaico_RR},  \ref{mosaicoSX}, \ref{SR} and \ref{mosaico_AC}, displaying data from 2000 and 2018 when available. The general data of the variable stars are listed in Table \ref{tab:variables}.

\begin{figure}
\centerline{\includegraphics[width=8.3cm]{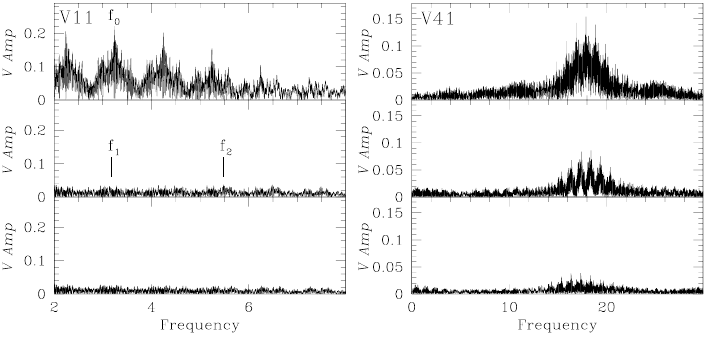}}
\caption{Frequency spectra of V11 and V41. In V11 no secondary frequencies are detected. The frequencies $f_1$ and $f_2$ indicate the locations where the secondary frequency found by \citet{Kopacki2007} and the second overtone would be found, respectively. In V41 three active frequencies are seen; the secondary frequencies remain after the consecutive prewhitening of the stronger signals.}
\label{V1141spec}
\end{figure}

\subsection{Multimode variables}
\subsubsection{RR Lyrae stars}

V11. Light curve shape variations were detected by \citet{Kopacki2001} who found two active frequencies, $f_0 = 3.2421$ d$^{-1}$ and $f_1 = 3.1715$ d$^{-1}$, interpreted respectively as the fundamental mode and a non-radial mode. The light curve in our observations is shown in Fig. \ref{mosaico_RR}. Clear amplitude and phase modulations are seen in the 2018 data (black symbols) and the light curve from 2000-2001 shows a remarkably larger amplitude. A period analysis of our 2000-2001 and 2018 data enabled to find the frequency $f_0 = 3.2420$ d$^{-1}$, in agreement with \citet{Kopacki2001} but, after prewhitening this value, no significant signal remains in the frequency spectrum of Fig. \ref{V1141spec}. In the figure, $f_1$ and $f_2$ mark the positions where the secondary frequency found by \citet{Kopacki2001} and the second overtone (5.4779 d$^{-1}$) would be found, respectively. No traces of $f_1$ are seen and $f_2$ is barely noticeable. Hence, we have to conclude that the stars is a monoperiodic RRc and that the evident amplitude modulations are most likely due to the Blazhko effect, well documented in many RRc-type stars (e.g. \citealt{Arellano2012}).

%\begin{figure*} 
%\includegraphics[width=17.0cm]{V11.pdf}
%\caption{Light curve of V11 phased with a model of 2 periods: P$_0$=0.308454 d and P$_1$=0.182553 d, the ratio P$_1$/P$_0$ =0.592 indicates that V11 pulsating in the fundamental and second-overtone modes. The left panel shows the data of \citet{Marin2002} and the right panel shows the 2018 data-set.}
%    \label{V11_DM}
%\end{figure*}

\begin{figure} 
\includegraphics[width=8.0cm]{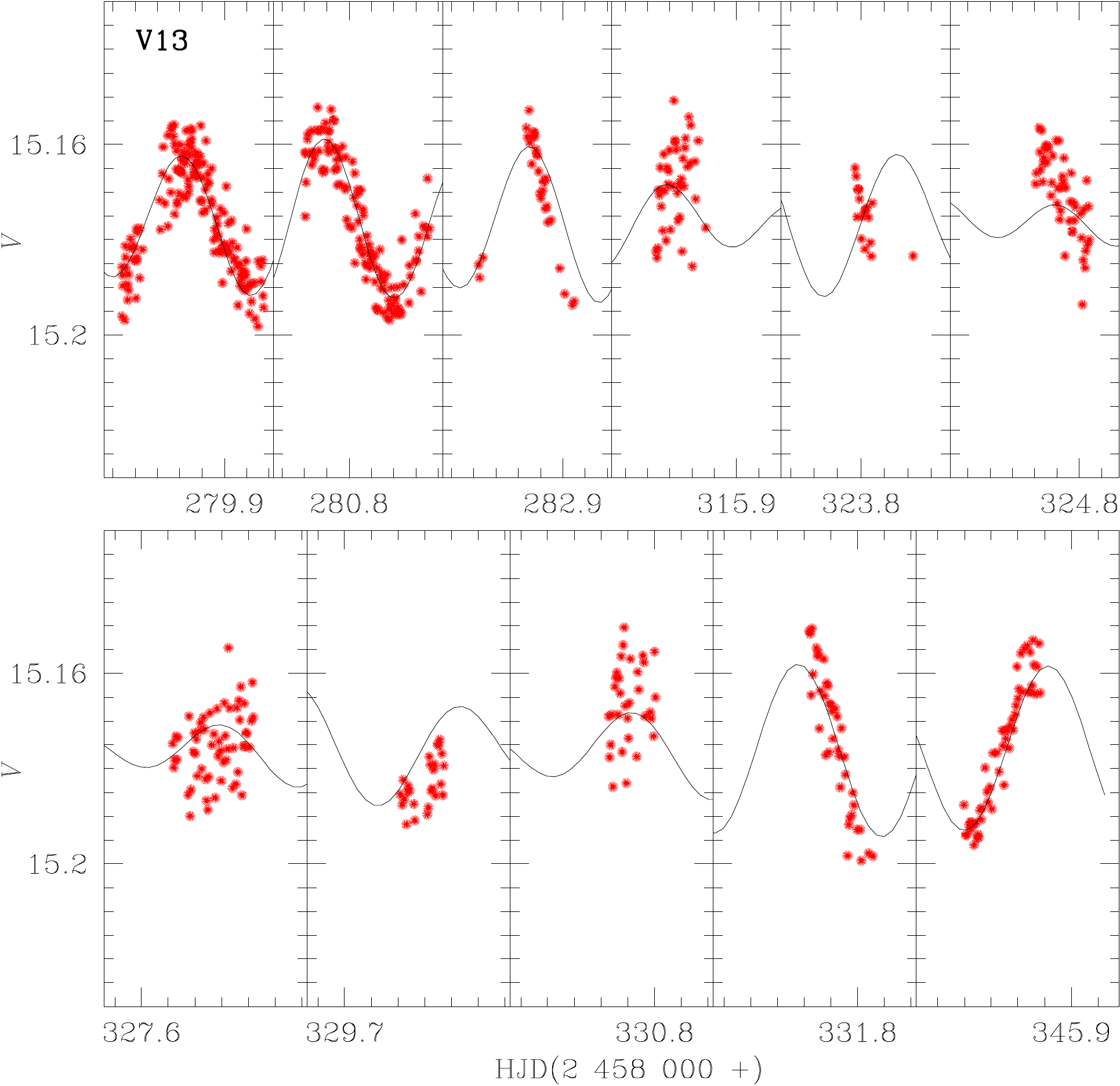}
\caption{Light curve of V13 with a two-period model fit with P$_0$=0.301536 d and P$_1$=0.337182 d is shown. The period ratio P$_1$/P$_0$= 0.89 suggests that at least one of the modes is non-radial.}
    \label{V13}
\end{figure}

V13. \citet{Hachenberg1939} could not classify this star, and although small amplitude variations were detected, \citet{Walker1955} was unable to determine its periodicity. \citet{Kopacki2001} found a period of  0.312955 d for this small amplitude RRc star. Its low amplitude is consistent with its very short period for a RRc star. In our observations we detected phase and amplitude modulations (Fig. \ref{mosaico_RR}) likely due to the presence of a second pulsating mode. The period analysis revealed the presence of two periods 0.301536 d and 0.337182 d with a ratio of 0.89, which suggests a double mode with at least one of them being non-radial. In Fig. \ref{V13} we show the comparison of the two mode model with the observations. The fit is generally satisfactory except in a few nights were the light curve is dominated by the scatter. Being a bright and isolated star, we rule out that these peculiarities are due to flux contamination. V13 is the bluest (hottest) RRc star in the cluster.

\begin{figure} 
\includegraphics[width=8.0cm]{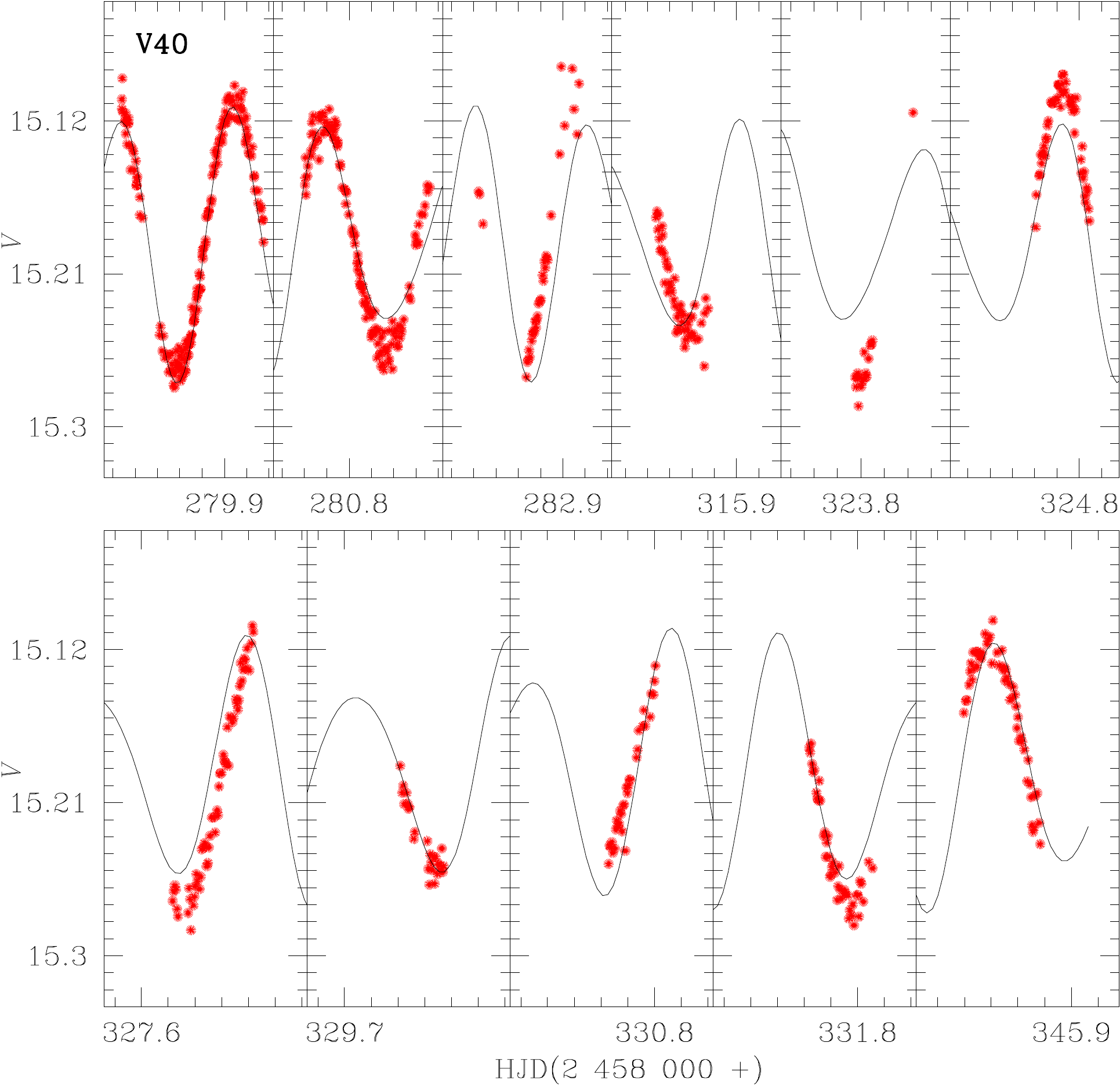}
\caption{Light curve of the RRd star V40. The two-period model fit with P$_0$=0.276639 d and $P_1$=0.201314 d is shown. The period ratio P$_1$/P$_0$= 0.73 suggests that both pulsations modes are radial.}
    \label{V40}
\end{figure}

V40.  This is a variable newly detected in the present work. Its period, light curve shape and position on the CMD, suggest that the star is of the RR Lyrae type. Its light curve in Fig.  \ref{mosaico_RR} displays amplitude modulations typical of double-mode pulsators. In fact, we have identified two active periods; $P_0=0.276639$ d and $P_1=0.201314$ d, for a ratio $P_1/P_0=0.73$ which suggest, two radial modes. Hence, the star is a typical double-mode or RRd star. The two-mode model and the fit to the data is displayed in Fig. \ref{V40}.

\begin{figure} 
\includegraphics[width=8.0cm]{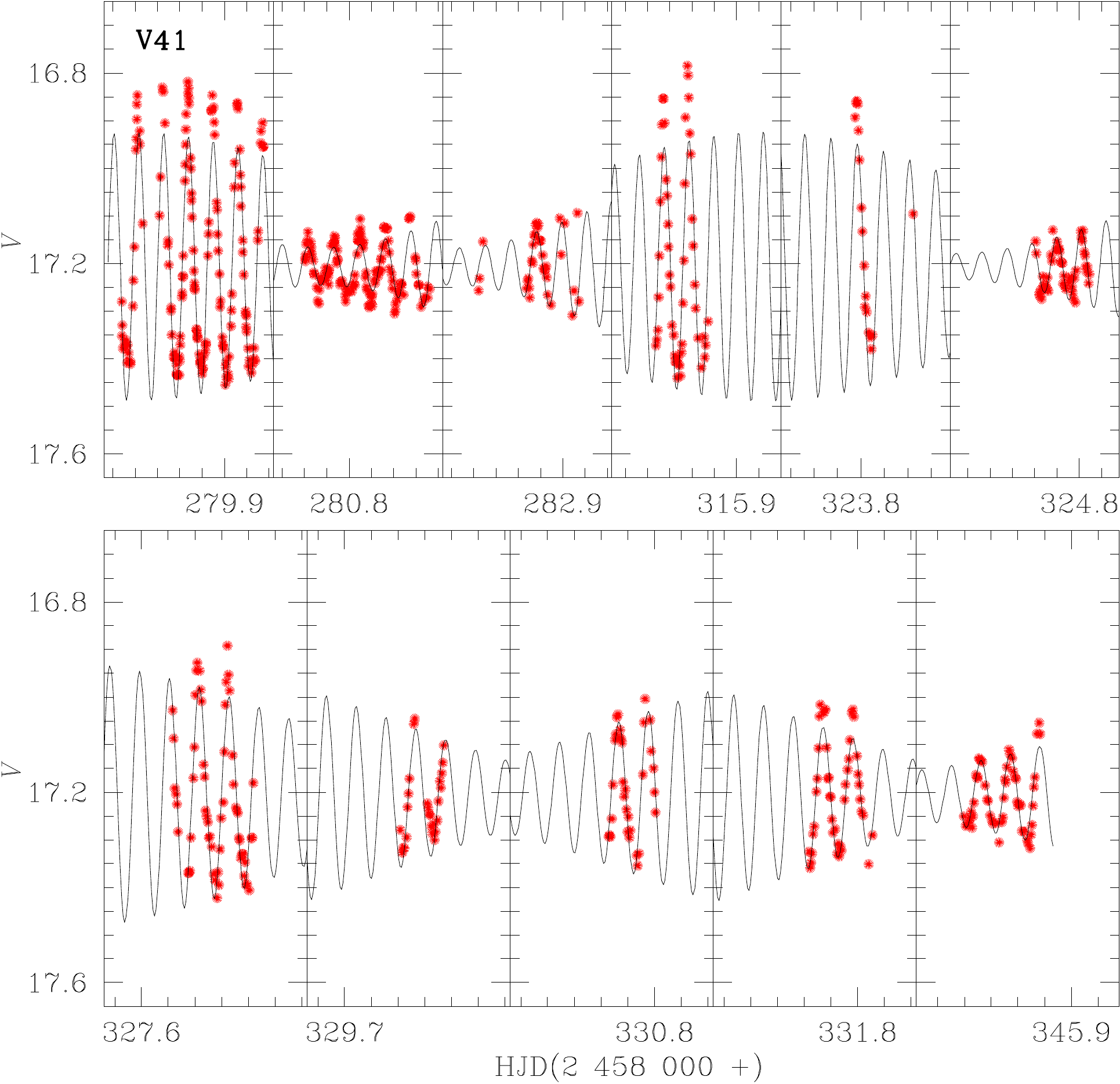}
\caption{Light curve of V41 phased with a three-period model: P$_0$=0.055947 d, P$_1$=0.054350 d, P$_2$=0.057632 d. According to the P-L relationship (Fig. \ref{SX_PL}), the period P$_0$ corresponds to fundamental mode.}
\label{V41}
\end{figure}

\subsubsection{SX Phe stars}

V41. This short period variable is a new detection in this work. On the CMD it is located in the blue straggler region and its light curve (Fig. \ref{mosaicoSX}) clearly indicates the presence of more that one period. We have classified it as SX Phe. According to the power spectrum obtained using Period04 and after successive prewhitenings of the main frequencies (see Fig. \ref{V1141spec}), we  could identify three active frequencies in our data: 17.873917 d$^{-1}$, 18.399359 d$^{-1}$, 17.351397 d$^{-1}$; or periods $P_0=0.055947$ d, $P_1=0.054350$ d, and $P_2=0.057632$ d, for the ratios $P_1/P_0= 0.97$ and  $P_2/P_0= 1.03$.
The fit to the data by the three-period model is shown in Fig. \ref{V41}.

\section{RR Lyrae light curve Fourier decomposition}
\label{FOURIER}

The RR Lyrae light curve morphology carries information on some of the fundamental physical parameters of the star, and it can be extracted by means of a proper Fourier decomposition and the use of $ad~hoc$ empirical or semi-empirical calibrations. This approach is particularly attractive to use on light curves of high quality and minimum scatter. To this end, the standard procedure is to represent the $V$ light curves by the Fourier series of harmonics of the form:

\begin{equation}
m(t) = A_0 ~+~ \sum_{k=1}^{N}{A_k ~\cos~\left( {2\pi \over P}~k~(t-E_0) ~+~ \phi_k \right) },
\label{eq_foufit}
\end{equation}

\noindent
where $m(t)$ is the magnitude at time $t$, $P$ is the period and $E_0$ the epoch, generally a time of maximum light. A linear minimization routine is used to derive the amplitudes $A_k$ and phases $\phi_k$ of each harmonic, from which the Fourier parameters $\phi_{ij} = j\phi_{i} - i\phi_{j}$ and $R_{ij} = A_{i}/A_{j}$ are calculated. The resulting intensity-weighted mean magnitudes $A_0 = <V>$, and the Fourier coefficients for the RRab and RRc stars are listed in Table \ref{tab:fourier_coeffs}. 

For the conversion of the Fourier parameters into [Fe/H], we used the calibrations of  \citet{Nemec2013}  based on new spectroscopic determinations of the metallicity, these calibrations are of the form; 
for RRab stars:
\begin{equation}
\label{NemRRab}
\begin{array}{cl}
{\rm [Fe/H]_{spec}}=&-8.65-40.12P+5.96\phi_{31}^s+6.27\phi_{31}^sP\\
&-0.72(\phi_{31}^s)^2,
\end{array}
\end{equation}
and for RRc stars:
\begin{equation}
\label{NemRRc}
\begin{array}{cl}
{\rm [Fe/H]_{spec}}=&1.70-15.67P+0.20\phi_{31}^c-2.41\phi_{31}^cP\\
&+18.0P^2+0.17(\phi_{31}^c)^2.   
\end{array}
\end{equation}

For completeness we have transformed from this spectroscopic scale back into the Zinn-West scale \citep{Zinn1984} (ZW) scale using the equation [Fe/H]$_{\rm spec} = -0.413 + 0.130$[Fe/H]$_{\rm ZW} -0.356$[Fe/H]$^2_{\rm ZW}$ \citep{Carretta2009}

In Table \ref{tab:fourier_coeffs} the Fourier coefficients of RRab and RRc stars are listed. The resulting physical parameters are reported in Table \ref{Tab:ParFis}. We do not include V29 and F1 in the weighted means of the physical parameters since these stars yield very different values of the distance. Their proper motions also indicate that these stars do not belong to the cluster (see Appendix \ref{IndStars}).

For the values of the absolute magnitude $M_V$, we used the calibrations of \citet{Kovacs2001} for the RRab stars and of \citet{Kovacs1998} for RRc stars. These calibrations are given in eqs. \ref{eq:KW01} and \ref{eq:K98} respectively.

\begin{equation} 
M_V = ~-1.876~\log~P ~-1.158~A_1 ~+0.821~A_3 + K.
\label{eq:KW01}
\end{equation} 

\begin{equation}
M_V = 1.061 ~-~ 0.961~P ~-~ 0.044~\phi^{(s)}_{21} ~-~ 4.447~A_4.
\label{eq:K98}	
\end{equation}

The above calibrations have an associated uncertainty of  0.04 mag. The zero points of the above equations; 1.061 and $K=0.41$ have been discussed by  \citet{Arellano2010}. 

The values of $M_{V}$ reported in Table \ref{Tab:ParFis} have been transformed to luminosities using the following equation:

\begin{equation}
 \log(L/L_{\odot}) = -0.4(M_{V} - M^{\odot}_{bol} + BC).
\end{equation}

To calculate the bolometric correction, we used the equation $BC = 0.06[\rm Fe/H]_{ZW} + 0.06$ as derived by \citet{SanCac1990}. We have adopted the value of  $M^{\odot}_{bol}$ = 4.75 mag.

For the temperature of the RRab and RRc stars we have followed the procedure described by \citet[their $\S$ 4.3]{Arellano2010} in which
the observed $(V-I)$  are converted into effective temperatures, via the polinomial calibration of \citet{Nemec2004}, which in turn uses the HB models of \citet{VandenBerg2006} and the colour–log $T_{\rm eff}$ relations as described
by \citet{VandenBerg2003}. The polinomial fit is of the form:

\begin{equation}
y = A_0 + A_1x + A_2x^2 + A_3x^3 + A_4x^4 + A_5x^5 + A_6x^6 + A_7x^7,
\end{equation}

\noindent
where y = log $T_{\rm eff}$ , x = $(V-I)_0$ and the coefficients $A_0 = 3.9867$,
$A_1 =-0.9506$, $A_2 = + 3.5541$, $A_3 =-3.4537$, $A_4 =-26.4992$,
$A_5 =+90.9507$, $A_6 =-109.6680$ and $A_7 =+46.7704$. The magnitude weighted $<V>$ - $<I>$ was dereddened assumnig $E(V-I) = 1.259 E(B-V)$.

We can also estimate the masses of the RR Lyrae stars using: 
$\rm log({\it M/M}_{\odot}) = 16.907 - 1.47\rm log(P_{F}) + 1.24\rm log({\it L/L}_{\odot}) - 5.12\rm log({\it T}_{\rm eff})$
as given by \citet{vanAlbada1971} where P$\rm_{F}$ is the fundamental period. The stellar mean radii can be estimated from $L  = 4\pi R^{2} \sigma T^{4}$. These values are also reported in Table \ref{Tab:ParFis}. \\ 

\begin{table*}
%\scriptsize
\begin{center}

\caption{Fourier coefficients of $V$ light curves of RRab and RRc stars in M92. The numbers in parentheses indicate the uncertainty on the last decimal place and represent the internal errors from the Fourier fit.} 

\label{tab:fourier_coeffs} 
\begin{tabular}{lllllllll}
\hline
ID  & $A_{0}$   & $A_{1}$  & $A_{2}$  & $A_{3}$  & $A_{4}$  &$\phi_{21}$&$\phi_{31}$&$\phi_{41}$  \\
\hline
\multicolumn{9}{c}{RRab} \\
%&&&&&RRab&&&&&\\
\hline
V1  & 15.109(1) & 0.312(1) & 0.156(1) & 0.100(1) & 0.063(1) & 4.016(10) & 8.379(15) & 6.594(22)  \\
V2  & 15.163(1) & 0.353(1) & 0.145(1) & 0.111(1) & 0.068(1) & 3.949( 9) & 7.985(12) & 6.039(18)  \\
V3  & 15.147(5) & 0.413(7) & 0.182(7) & 0.139(7) & 0.108(7) & 3.931(50) & 8.022(71) & 5.973(91)  \\
V4  & 15.073(1) & 0.339(2) & 0.160(2) & 0.121(2) & 0.073(1) & 3.803(13) & 7.982(17) & 5.940(27)  \\
V5  & 15.193(1) & 0.357(1) & 0.153(1) & 0.123(1) & 0.081(1) & 3.843( 9) & 7.967(13) & 5.911(18)  \\
V6  & 15.122(1) & 0.430(2) & 0.192(2) & 0.124(2) & 0.075(2) & 3.882(14) & 8.083(21) & 5.991(32)  \\
V8  & 15.179(1) & 0.328(1) & 0.152(1) & 0.103(1) & 0.069(1) & 3.988(13) & 8.169(19) & 6.179(27)  \\
V9  & 15.198(2) & 0.423(4) & 0.163(3) & 0.085(3) & 0.049(3) & 3.847(24) & 7.832(45) & 5.052(68)  \\
V25 & 15.131(1) & 0.277(1) & 0.131(1) & 0.087(1) & 0.055(1) & 4.024(14) & 8.359(21) & 6.518(31)  \\
V29 & 15.425(1) & 0.442(2) & 0.174(2) & 0.151(2) & 0.103(2) & 3.845(12) & 7.875(16) & 5.881(23) \\
\hline
\multicolumn{9}{c}{RRc} \\
%&&&&&RRc&&&&&\\
\hline
%var   A0   s   A1  s   A2  s   A3  s   A4  s fi21    s  fi31   s  fi41  s     P   
V10 & 15.187(1) & 0.264(1) & 0.037(1) & 0.022(1) & 0.011(1) & 4.719( 20) & 3.324( 33) & 2.190( 66)  \\
V12 & 15.081(1) & 0.217(1) & 0.021(1) & 0.014(1) & 0.005(1) & 5.110( 32) & 3.716( 47) & 2.866(117)  \\
V31 & 14.880(2) & 0.185(2) & 0.028(2) & 0.017(2) & 0.012(2) & 5.256( 82) & 3.685(130) & 2.365(185)  \\
V32 & 15.166(1) & 0.110(1) & 0.007(1) & 0.001(1) & 0.001(1) & 4.873( 70) & 2.764(363) & 1.887(386)  \\
F1 & 14.697(1) & 0.020(1) & 0.002(1) & 0.002(1) & 0.001(1) & 5.499(290) & 0.752(317) & 2.146(970)  \\
\hline
\multicolumn{9}{c}{AC} \\
%&&&&&RRc&&&&&\\
\hline
%var   A0   s   A1  s   A2  s   A3  s   A4  s fi21    s  fi31   s  fi41  s     P   
V7  & 14.207(1) & 0.254(3) & 0.119(4) & 0.073(2) & 0.033(2) & 3.963(34) & 8.052(75) & 6.397(111)  \\
\hline
\multicolumn{9}{c}{BL Her} \\
%&&&&&RRc&&&&&\\
\hline
V30 & 13.947(1) & 0.221(2) & 0.072(1) & 0.022(1) & 0.006(1) & 3.754(24) & 8.182(63) & 4.885(24)  \\
\hline
\end{tabular}
\end{center}
\end{table*}

\begin{table*}
%\scriptsize
\begin{center}

\caption{Physical parameters of the RRab and RRc stars in M92. The numbers in parentheses 
indicate the internal errors on the last decimal place; $\sigma$ is the standard deviation of the mean.} 
\label{Tab:ParFis}

\begin{tabular}{cllllllll}
\hline
\multicolumn{9}{c}{RRab} \\
\hline 
ID&[Fe/H]$_{\rm ZW}$ & [Fe/H]$_{\rm spec}$ &$M_V$ & log~$T_{\rm eff}$  &log$(L/{
L_{\odot}})$ &$M/{ M_{\odot}}$&$R/{ R_{\odot}}$ & $D(kpc)$\\
\hline
\hline
V1  & -1.86( 3) & -1.89( 4) & 0.41( 1) & 3.797 & 1.736(1) & 0.70 & 6.32 & 8.46 \\
V2  & -2.25( 2) & -2.51( 3) & 0.45( 1) & 3.792 & 1.720(1) & 0.80 & 6.33 & 8.51 \\
V3  & -2.16(12) & -2.36(19) & 0.41(10) & 3.792 & 1.735(4) & 0.85 & 6.44 & 8.60 \\
V4  & -2.18( 3) & -2.39( 5) & 0.49( 3) & 3.794 & 1.702(1) & 0.78 & 6.17 & 8.01 \\
V5  & -2.16( 2) & -2.35( 4) & 0.48( 1) & 3.799 & 1.706(1) & 0.76 & 6.04 & 8.50 \\
V6  & -1.86( 4) & -1.89( 5) & 0.43( 3) & 3.809 & 1.728(1) & 0.75 & 5.91 & 8.43 \\
V8  & -2.09( 3) & -2.25( 5) & 0.44( 1) & 3.793 & 1.725(1) & 0.76 & 6.34 & 8.63 \\
V9 & -2.32(7) & -2.62(13) & 0.39(1) & 3.799 & 1.743(2) & 0.85 & 6.29 & 8.88 \\
V25 & -1.89( 4) & -1.93( 6) & 0.45( 1) & 3.791 & 1.720(1) & 0.72 & 6.36 & 8.39 \\
V29$^{1}$ & -2.17( 3) & -2.375( 4) & 0.44( 3) & 3.799 & 1.723(1) & 1.02 & 6.66 & 9.64 \\
\hline
Weighted mean &-2.10(1) &-2.22(2) &0.45(1) &3.796 &1.721(1) &0.77 &6.24 &8.49 \\
$\sigma$ &$\pm$0.17 &$\pm$0.28 &$\pm$0.03 &$\pm$0.006 &$\pm$0.014 &$\pm$0.05 &$\pm$0.17 &$\pm$ 0.23 \\
\hline
\multicolumn{9}{c}{RRc} \\
\hline
ID&[Fe/H]$_{\rm ZW}$ & [Fe/H]$_{\rm spec}$ &$M_V$ & log~$T_{\rm eff}$  &log$(L/{
L_{\odot}})$ &$M/{ M_{\odot}}$&$R/{R_{\odot}}$& $D(kpc)$\\
\hline

\hline
V10 & -2.02( 1) & -2.13(1) & 0.51(1) & 3.823 & 1.696(2) & 0.75 & 5.35 & 8.37 \\
V12 & -2.11( 2) & -2.28(1) & 0.49(1) & 3.816 & 1.704(2) & 0.73 & 5.58 & 8.05 \\
V31$^{1}$ & -2.05( 6) & -2.18(4) & 0.46(1) & 3.854 & 1.715(4) & 0.51 & 4.75 & 7.43 \\
V32 & -1.80(13) & -1.80(9) & 0.60(1) & 3.817 & 1.660(2) & 0.90 & 5.27 & 7.96 \\
F1$^{1}$ & -1.86(11) & -1.89(8) & 0.56(1) & 3.863 & 1.675(5) & 0.52 & 4.34 & 6.52 \\
\hline
Weighted mean &-2.04(1) &-2.17(1) &0.53(1) &3.819  &1.687(1) & 0.79 &5.40 &8.13 \\
$\sigma$ &$\pm$0.16 &$\pm$0.24 &$\pm$0.06 &$\pm$0.004 &$\pm$0.023 &$\pm$0.09 &$\pm$0.16 &$\pm$0.22\\
\hline
\end{tabular}
\center{1. Not included in the average of the physical parameters. F1 is not a cluster member.}
\end{center}
\end{table*}

\subsection{Bailey diagram and Oosterhoff type}
\label{secBailey}

The period versus amplitude plane for RR Lyrae is known as the Bailey Diagram. It's very useful for separating RRab from RRc stars. This diagram offers insight on the Oosterhoff type of a globular cluster i.e. Oo I or Oo II. 

\begin{figure}
\includegraphics[width=8cm]{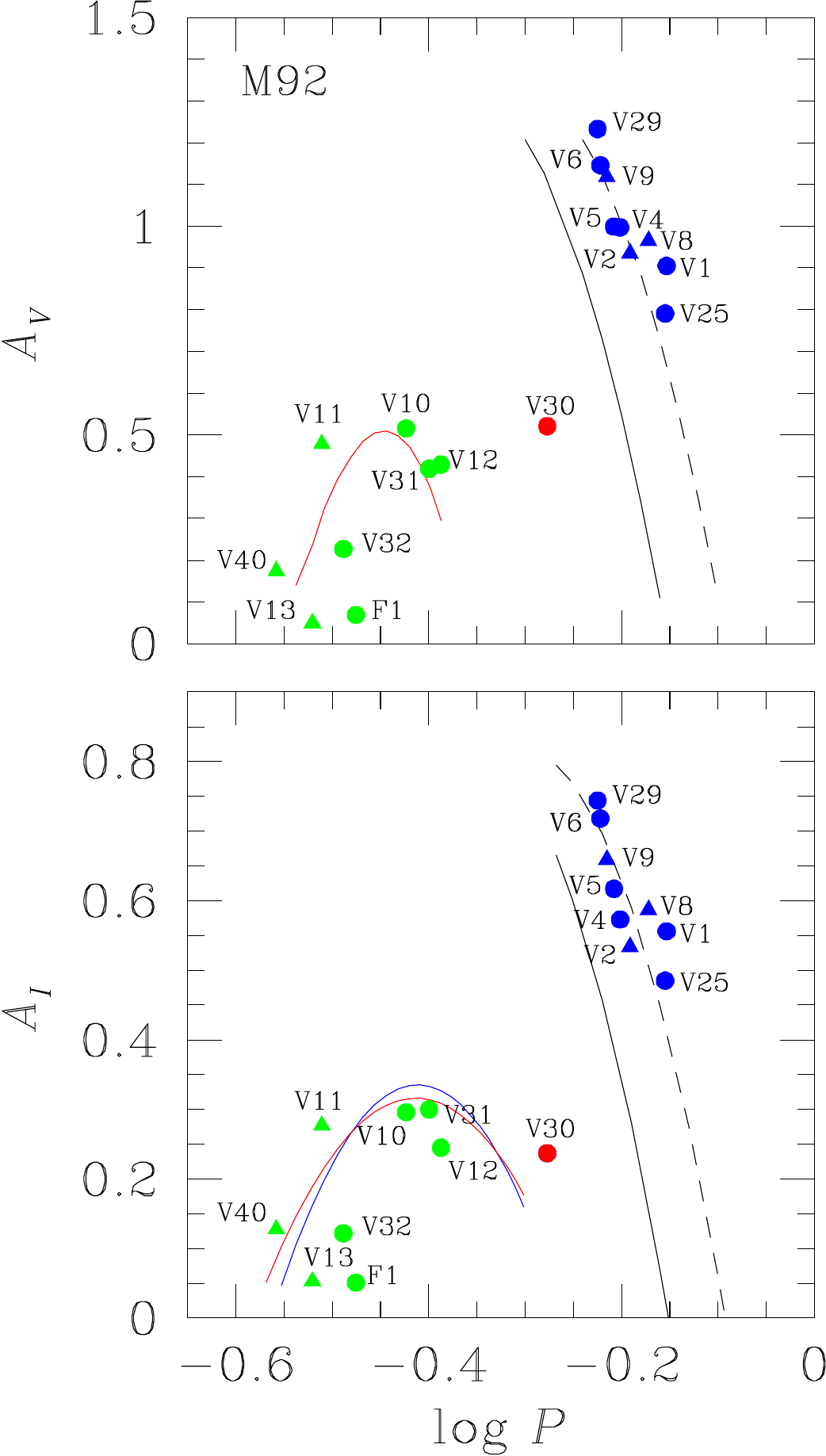}
\caption{Bailey diagram for M92. Blue and green symbols represent RRab and RRc stars, respectively. Triangles correspond to stars with amplitude modulations. The continuous and segmented lines are the calibrations described in $\S$ \ref{secBailey}. The distribution of RRab star clearly show the Oo II nature of M92. The star V30 is definitely not a RRab star but instead a BL Her-type star.}
\label{figBailey} 
\end{figure}

Fig. \ref{figBailey} shows the $V$ and $I$ amplitudes of the RR Lyrae stars as a function of their period. In the top panel are the loci for unevolved and evolved stars in M3 according to \citet{Cacciari2005}. The red parabola was obtained by \citet{Kunder2013a} for RRc stars in 14 OoII clusters. In the bottom panel the black solid and segmented loci for unevolved and evolved stars respectively are from \citet{Kunder2013b}. The red parabola was obtained by \citet{Deras2019}. 
We extended this later calibration, by using Period-Amplitude data for RRc stars in seven Oo II type globular clusters (including our M92 data), for  a total of 28 RRc stars in NGC 288, NGC 6333, NGC 1904, NGC 7099, M53 and M13.

The corresponding equation for our calibration is:

\begin{equation}
\begin{array}{cl}
A_I=& (-11.00 \pm 4.70) (\log {\rm P} )^2 -(9.10 \pm 4.18) \log {\rm P}\\
&- (1.56 \pm 0.92)
\end{array}
\end{equation}

\begin{figure}
\includegraphics[width=8cm]{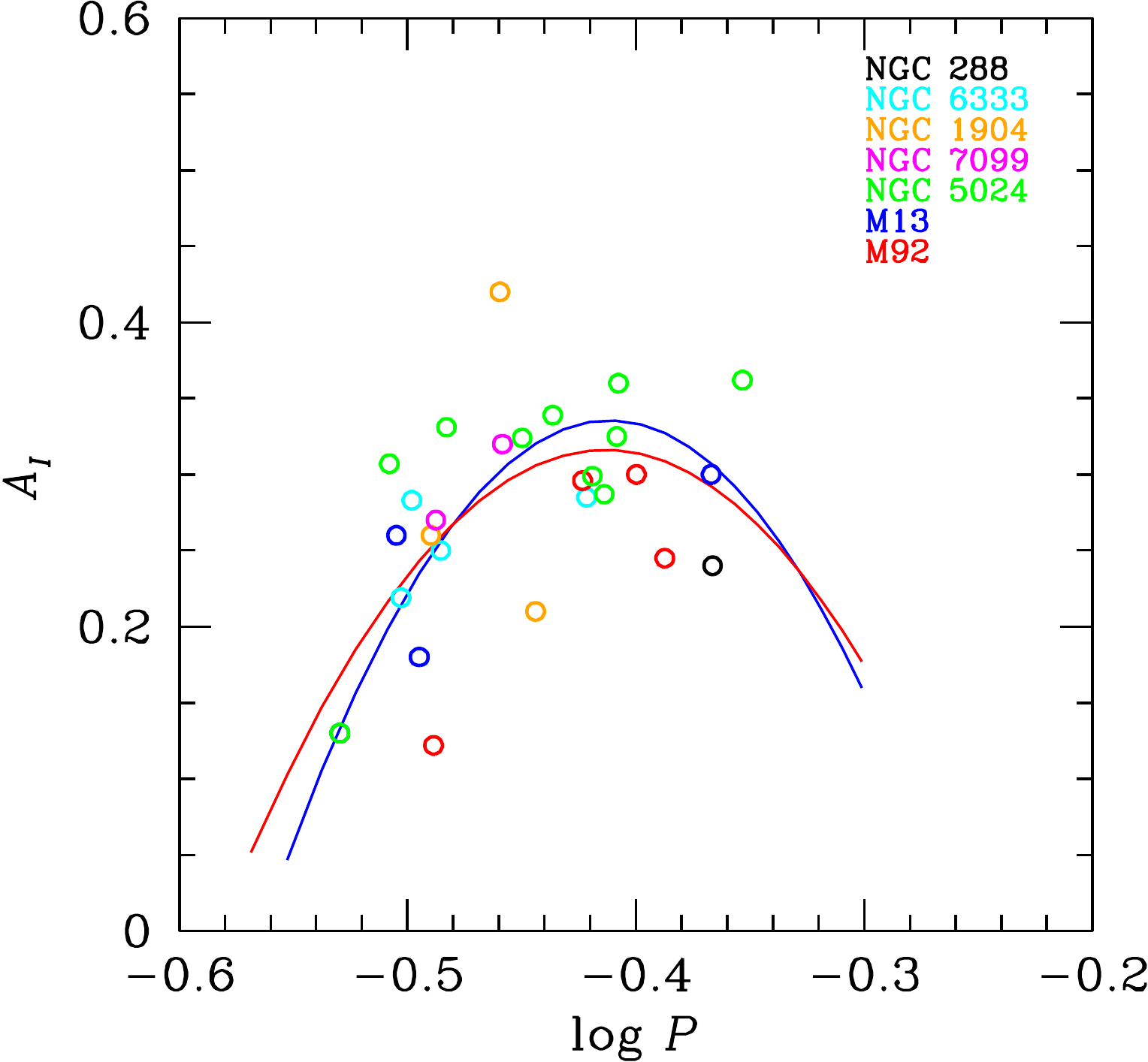}
\caption{Distribution of RRc stars in the period-amplitude plane for Oo II type clusters. The red parabola was calculated by \citet{Deras2019} for RRc stars in M13. The blue parabola was calculated in this work using 28 RRc stars from 7 Oo~II type globular clusters.}
\label{fitBailey} 
\end{figure}

We can see that this parabola fits well even those stars that present amplitude modulations. The diagram clearly shows the Oo II type nature of M92. The red and blue parabolas in Fig. \ref{fitBailey} are the calibrations of \citet{Deras2019} and our own, respectively.

\subsection{Notes on the SX Phe distance}

An approach to the calculation of the cluster distance was made by means of the the Period-Luminosity relationship for SX Phe stars. A calibration of the fundamental mode period was calculated by  \citet{CohenSara2012}, and it is of the form:

\begin{equation}
M_V=-(3.389\pm0.090)\log {\rm P} -(1.640\pm0.110).
\label{eq:SX_PL}
\end{equation}

\begin{figure} 
\includegraphics[width=8.0cm]{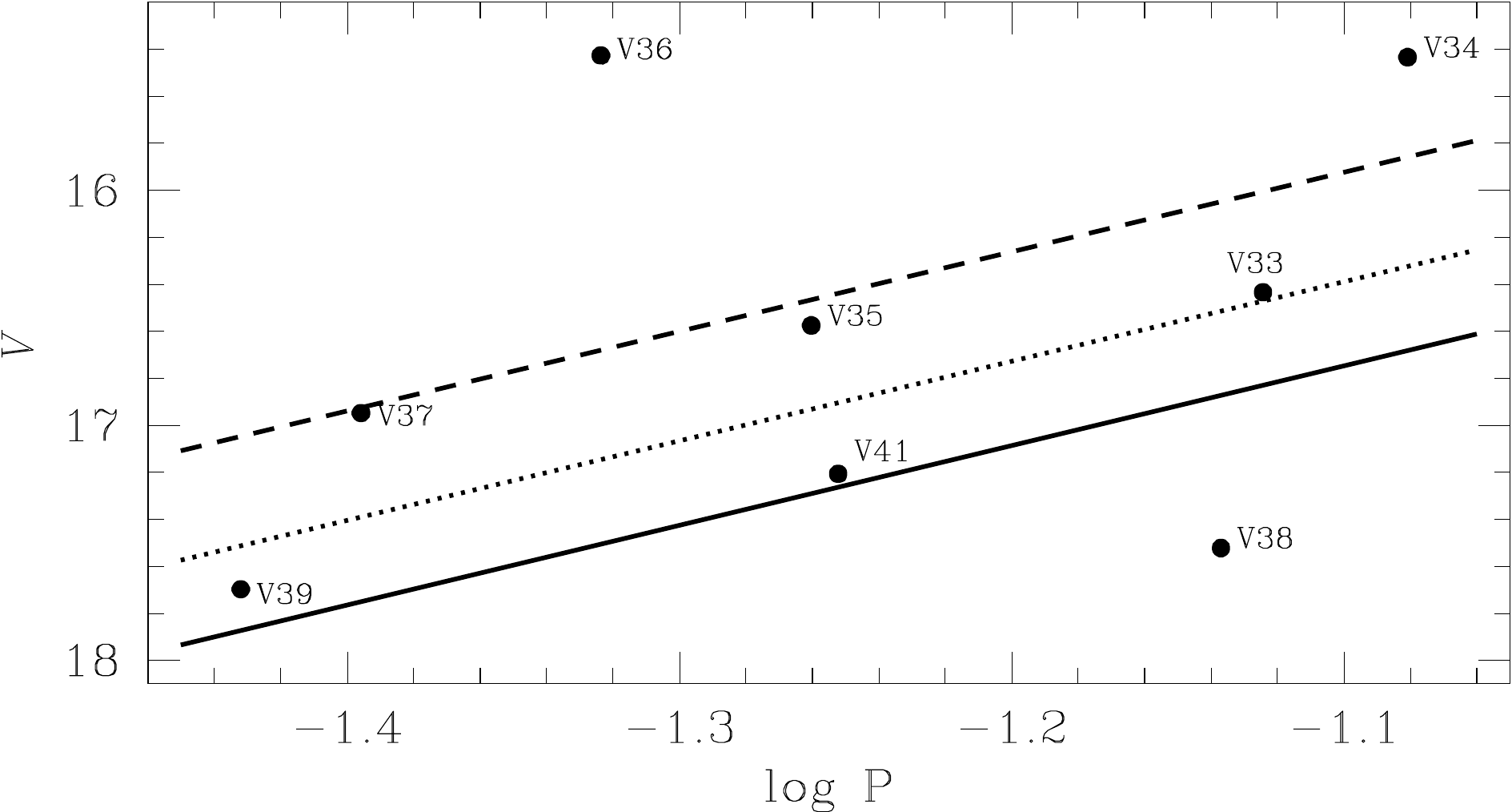}
\caption{P-L relationship for the SX Phe stars. The solid line represents the P-L calibration of \citet{CohenSara2012} placed at a distance of $8.31\pm0.16$ kpc (mean of the value derived from RR Lyrae stars results) and $E(B-V)$=0.02. The dotted and dashed lines correspond to the first and second overtones, respectively.}
    \label{SX_PL}
\end{figure}

This calibration has an associated uncertainty of 0.104 mag. Fig. \ref{SX_PL} shows this relationship positioned at a distance of $8.31\pm0.16$ kpc (mean of the distance derived from the RR Lyrae stars Fourier decomposition) and $E(B-V)$=0.02. The solid line corresponds to fundamental mode, the dotted and dashed lines to the first and second overtones respectively, assuming $P_1/P_0=0.783$ and $P_2/P_0=0.571$. Plotting the 8 SX Phe stars in the field of M92, the figure suggests that V41 pulsates in the fundamental mode,  V39 is half-way between the fundamental mode and the first overtone; V33 pulsates in the first
and V35 and V37 in the second overtone, however, as discussed in Appendix \ref{IndStars}, their magnitudes are contaminated by the presence of a close neighbour, hence their positions are most likely spurious and were not included in the distance estimation.
V34, V36 and V38 lie away from the relationship trend, which is not surprising considering that these three stars were found not to be cluster members (Appendix \ref{IndStars}).

An application of eq. \ref{eq:SX_PL} for V35, V39 (fundamental) and V42 leads to a mean distance of $8.2\pm0.4$ kpc, in excellent agreement with the distance found from the Fourier decomposition of the RR Lyrae stars. If V39 is assumed to be in the first overtone the distance would be $8.6\pm0.4$ kpc.

\subsection{Comparison of [Fe/H] and distance with literature values}

In Table \ref{Tab:Literature} we summarise previous estimates of metallicity, cluster mean distance, reddening and age for M92 found in the literature and compare the metallicity and distance values with our results. Our values depict the cluster to be slightly more metal-rich and closer than most determinations but within the respective uncertainties the match is good.

The values of reddening and age estimations are also included for completeness. For the reddening we have simply adopted $E(B-V)=0.02$. Most of the age calculations describe M92 as a very old cluster with an age between 12.5 and 13.2 Gyrs. In the CMD of Fig. \ref{CMD_M92} we show three isochrones with ages of 12, 12.5 and 13 Gyrs built from the Victoria-Regina models \citep{VandenBerg2014}, and note that the models between 12.5-13 Gyrs (blue and green isochrones in Fig. \ref{CMD_M92}) best match our observations. If we plot isochrones of the ages reported by \citet{Di_Cecco2010}, they would be not consistent with our data.

\begin{table}
\scriptsize
\begin{center}
\caption{Other determinations of metallicity, distance, reddening and age for M92.} 
\label{Tab:Literature}
\begin{tabular}{lcccc}
\hline 
Literature & [Fe/H] & Distance & E(B-V) & Age \\
           &        &  (kpc)   &        & (Gyrs)\\
\hline
\citet{Di_Cecco2010} & -2.32 & 8.25 & 0.025 & 11 $\pm$ 1.5 \\
CNO-enhanced models &       &       &       & 10 $\pm$ 1.5 \\
\hline
\citet{Dotter2010}   & -2.4  & 8.73$^1$ & 0.031$^1$ &  13.25 $\pm$ 1 \\
\hline
\citet{VandenBerg2014}& -2.4  & 8.75 & 0.019 & 13 \\
\hline
\citet{VandenBerg2016}& -2.3  & 8.59 & 0.023 & 12.5-12.9 \\
                     & -2.6  & 8.75 & 0.023 & 12.8 \\
\hline
This work &        &      & 0.02$^2$ &  \\
RRab      & -2.22$\pm$0.28 & 8.49$\pm$ 0.23 &  & -- \\
RRc       & -2.17$\pm$0.24 & 8.13$\pm$ 0.22 &  & \\
\hline
\end{tabular}
\center{1. These values are given for (m-M)$_{F814W}$ and E(F606W-F814W). 2. Adopted value.}
\end{center}
\end{table}

\section{Summary and conclusions}
\label{conclusion}

Time series \emph{VI} CCD photometry of a $7.5\times7.5$ arcmin$^2$ FoV around M92 allowed us to study most of the known variable stars in the cluster. We were able to identify four new variables V40-V42 and F1, of the types RRd, SX Phe, SR and RRc, respectively. We demonstrate that F1 is a field variable. Multiple modes were also identified in the RRc stars V13 and V40 and the SX Phe V41. The double mode nature of V11 is not confirmed but it is rather considered a Blazhko pulsator with clear amplitude and phase modulations. We confirm V7 as AC and argue that V30, also located more than 1.0 mag above the HB, is a non-cluster member BL Her-type star.

Using the proper motions from $Gaia$-DR2, the method of \citet{Bustos_Fierro2019}, 5012 stars were identified as likely cluster members, out of the 6974 stars measured by our photometric pipeline DanDIA. This enabled us to produce a cleaner CMD. 
The position of the variables on the CMD and their proper motions allowed discussion of their membership to the cluster. We found that V29, V30, V31, V34, V36 and the newly discovered RRc F1, do not belong to the cluster, as is not the variable V14 (not in our field) \citep{Rees1992}.

Via the Fourier decomposition of the RR Lyrae light curves and the calibrations of \citet{Nemec2013} for the iron abundance, we calculated a mean value of [Fe/H]$_{\rm spec} = -2.20 \pm 0.18$ which is given in the spectroscopic scale, and a mean distance to the cluster of $8.3\pm0.2$ kpc. These values are in good concordance with previous estimations from independent methods.

M92 is a clear Oo II-type cluster, with numerous indicators: The mean period of the RRab stars in M92 is $\left<P_{ab}\right>$=0.65 d. Its period-amplitude, or Bailey, diagram shows a neat distribution of the RRab stars along the evolved star sequence identified in M3 \citep{Jurcsik2015}. It is among the Galactic clusters with the bluest HB and it is extremely metal-poor. The distribution of RRab and RRc stars on the HB displays a clear segregation of modes, which seems to be the rule in all Oo II-type clusters, but not in all of the Oo I-type, where there are cases of occupation of both modes in the bi-modal region of the instability strip, or "either-or" region. \citep[see their Figure 8]{Arellano2019}. This result has its relevance in the fact that the distribution of pulsation modes on the HB is a consequence of the initial mass distribution of newcomers to the instability strip on the ZAHB \citep{Caputo1978}. This in turn, depends on the mass loss during the He-flashes at the RGB phase and on the pre-HB phase related to
the onset of the CNO cycle. This may imply that RR Lyrae stars in Oo II clusters are always evolved from less massive stars starting their HB evolution on the bluer part of the ZAHB, whereas Oo I-type clusters may have a wider distribution of masses on the ZAHB. What determines the two possibilities in Oo I clusters is not clear, but it is most likely connected with the mass loss during the RGB phase. For further details we refer the reader to the discussion by \citet[their section 7]{Yepez2018}.

Among the SX Phe stars in the field of M92, three were found not to belong to the cluster, namely V34, V36 and V38. Among the member SX Phe, two of them, V33 and V37, could not be resolved from a very close neighbour of similar brightness but, their proper motions indicate that they are likely cluster members. The remaining three V35, V39 and V41, through their P-L relation, suggest a mean distance of $8.2\pm0.4$ kpc, in good agreement with the distance found from the RR Lyrae stars.

\section*{ACKNOWLEDGEMENTS}
We are grateful to Drs. I. Bustos Fierro and J.H. Calder\'on for identifying the most probable cluster members in M92. This project was partially supported by DGAPA-UNAM (Mexico) via grant IG100620. MAY and DD thank CONACyT for the PhD scholarships. We have made extensive use of the SIMBAD and ADS services.

\bibliographystyle{mnras}
\bibliography{M92} % if your bibtex file is calledexample.bib

\begin{thebibliography}{}
\makeatletter
\relax
\def\mn@urlcharsother{\let\do\@makeother \do\$\do\&\do\#\do\^\do\_\do\%\do\~}
\def\mn@doi{\begingroup\mn@urlcharsother \@ifnextchar [ {\mn@doi@}
  {\mn@doi@[]}}
\def\mn@doi@[#1]#2{\def\@tempa{#1}\ifx\@tempa\@empty \href
  {http://dx.doi.org/#2} {doi:#2}\else \href {http://dx.doi.org/#2} {#1}\fi
  \endgroup}
\def\mn@eprint#1#2{\mn@eprint@#1:#2::\@nil}
\def\mn@eprint@arXiv#1{\href {http://arxiv.org/abs/#1} {{\tt arXiv:#1}}}
\def\mn@eprint@dblp#1{\href {http://dblp.uni-trier.de/rec/bibtex/#1.xml}
  {dblp:#1}}
\def\mn@eprint@#1:#2:#3:#4\@nil{\def\@tempa {#1}\def\@tempb {#2}\def\@tempc
  {#3}\ifx \@tempc \@empty \let \@tempc \@tempb \let \@tempb \@tempa \fi \ifx
  \@tempb \@empty \def\@tempb {arXiv}\fi \@ifundefined
  {mn@eprint@\@tempb}{\@tempb:\@tempc}{\expandafter \expandafter \csname
  mn@eprint@\@tempb\endcsname \expandafter{\@tempc}}}

\bibitem[\protect\citeauthoryear{{Arellano Ferro}, {Giridhar}  \&
  {Bramich}}{{Arellano Ferro} et~al.}{2010}]{Arellano2010}
{Arellano Ferro} A.,  {Giridhar} S.,   {Bramich} D.~M.,  2010, \mnras, \href
  {https://ui.adsabs.harvard.edu/\#abs/2010MNRAS.402..226A} {402, 226}

\bibitem[\protect\citeauthoryear{{Arellano Ferro}, {Bramich}, {Figuera Jaimes},
  {Giridhar}  \& {Kuppuswamy}}{{Arellano Ferro} et~al.}{2012}]{Arellano2012}
{Arellano Ferro} A.,  {Bramich} D.~M.,  {Figuera Jaimes} R.,  {Giridhar} S.,
  {Kuppuswamy} K.,  2012, \mn@doi [\mnras] {10.1111/j.1365-2966.2011.20119.x},
  \href {https://ui.adsabs.harvard.edu/abs/2012MNRAS.420.1333A} {420, 1333}

\bibitem[\protect\citeauthoryear{{Arellano Ferro}, {Luna}, {Bramich},
  {Giridhar}, {Ahumada}  \& {Muneer}}{{Arellano Ferro}
  et~al.}{2016}]{Arellano2016}
{Arellano Ferro} A.,  {Luna} A.,  {Bramich} D.~M.,  {Giridhar} S.,  {Ahumada}
  J.~A.,   {Muneer} S.,  2016, \apss, \href
  {http://adsabs.harvard.edu/abs/2016Ap%26SS.361..175A} {361, 175}

\bibitem[\protect\citeauthoryear{{Arellano Ferro}, {Bustos Fierro},
  {Calder{\'o}n}  \& {Ahumada}}{{Arellano Ferro} et~al.}{2019}]{Arellano2019}
{Arellano Ferro} A.,  {Bustos Fierro} I.~H.,  {Calder{\'o}n} J.~H.,   {Ahumada}
  J.~A.,  2019, \rmxaa, \href
  {https://ui.adsabs.harvard.edu/abs/2019RMxAA..55..337A} {55, 337}

\bibitem[\protect\citeauthoryear{{Baumgardt}, {Hilker}, {Sollima}  \&
  {Bellini}}{{Baumgardt} et~al.}{2019}]{Baumgardt2019}
{Baumgardt} H.,  {Hilker} M.,  {Sollima} A.,   {Bellini} A.,  2019, \mn@doi
  [\mnras] {10.1093/mnras/sty2997}, \href
  {https://ui.adsabs.harvard.edu/abs/2019MNRAS.482.5138B} {482, 5138}

\bibitem[\protect\citeauthoryear{{Bramich}}{{Bramich}}{2008}]{Bramich2008}
{Bramich} D.~M.,  2008, \mnras, \href
  {http://adsabs.harvard.edu/abs/2008MNRAS.386L..77B} {386, L77}

\bibitem[\protect\citeauthoryear{{Bramich}, {Figuera Jaimes}, {Giridhar}  \&
  {Arellano Ferro}}{{Bramich} et~al.}{2011}]{Bramich2011}
{Bramich} D.~M.,  {Figuera Jaimes} R.,  {Giridhar} S.,   {Arellano Ferro} A.,
  2011, MNRAS, \href {http://adsabs.harvard.edu/abs/2011MNRAS.413.1275B} {413,
  1275}

\bibitem[\protect\citeauthoryear{{Bramich} et~al.,}{{Bramich}
  et~al.}{2013}]{Bramich2013}
{Bramich} D.~M.,  et~al., 2013, \mnras, \href
  {http://adsabs.harvard.edu/abs/2013MNRAS.428.2275B} {428, 2275}

\bibitem[\protect\citeauthoryear{{Burke}, {Rolland}  \& {Boy}}{{Burke}
  et~al.}{1970}]{Burke1970}
{Burke} Edward~W. J.,  {Rolland} W.~W.,   {Boy} W.~R.,  1970, \jrasc, \href
  {https://ui.adsabs.harvard.edu/abs/1970JRASC..64..353B} {64, 353}

\bibitem[\protect\citeauthoryear{{Bustos Fierro} \& {Calder{\'o}n}}{{Bustos
  Fierro} \& {Calder{\'o}n}}{2019}]{Bustos_Fierro2019}
{Bustos Fierro} I.~H.,  {Calder{\'o}n} J.~H.,  2019, \mnras, \href
  {https://ui.adsabs.harvard.edu/abs/2019MNRAS.488.3024B} {488, 3024}

\bibitem[\protect\citeauthoryear{{Cacciari}, {Corwin}  \& {Carney}}{{Cacciari}
  et~al.}{2005}]{Cacciari2005}
{Cacciari} C.,  {Corwin} T.~M.,   {Carney} B.~W.,  2005, \aj, \href
  {http://adsabs.harvard.edu/abs/2005AJ....129..267C} {129, 267}

\bibitem[\protect\citeauthoryear{{Caputo}, {Castellani}  \&
  {Tornambe}}{{Caputo} et~al.}{1978}]{Caputo1978}
{Caputo} F.,  {Castellani} V.,   {Tornambe} A.,  1978, \aap, \href
  {https://ui.adsabs.harvard.edu/#abs/1978A&A....67..107C} {67, 107}

\bibitem[\protect\citeauthoryear{{Carretta}, {Bragaglia}, {Gratton}, {D'Orazi}
  \& {Lucatello}}{{Carretta} et~al.}{2009}]{Carretta2009}
{Carretta} E.,  {Bragaglia} A.,  {Gratton} R.,  {D'Orazi} V.,   {Lucatello} S.,
   2009, \aap, \href {http://adsabs.harvard.edu/abs/2009A%26A...508..695C}
  {508, 695}

\bibitem[\protect\citeauthoryear{{Clement} et~al.,}{{Clement}
  et~al.}{2001}]{Clement2001}
{Clement} C.~M.,  et~al., 2001, \aj, \href
  {http://adsabs.harvard.edu/abs/2001AJ....122.2587C} {122, 2587}

\bibitem[\protect\citeauthoryear{{Cohen} \& {Sarajedini}}{{Cohen} \&
  {Sarajedini}}{2012}]{CohenSara2012}
{Cohen} R.~E.,  {Sarajedini} A.,  2012, \mnras, \href
  {http://adsabs.harvard.edu/abs/2012MNRAS.419..342C} {419, 342}

\bibitem[\protect\citeauthoryear{{Deras}, {Arellano Ferro}, {L{\'a}zaro},
  {Bustos Fierro}, {Calder{\'o}n}, {Muneer}  \& {Giridhar}}{{Deras}
  et~al.}{2019}]{Deras2019}
{Deras} D.,  {Arellano Ferro} A.,  {L{\'a}zaro} C.,  {Bustos Fierro} I.~H.,
  {Calder{\'o}n} J.~H.,  {Muneer} S.,   {Giridhar} S.,  2019, \mnras, \href
  {https://ui.adsabs.harvard.edu/abs/2019MNRAS.486.2791D} {486, 2791}

\bibitem[\protect\citeauthoryear{{Di Cecco} et~al.,}{{Di Cecco}
  et~al.}{2010}]{Di_Cecco2010}
{Di Cecco} A.,  et~al., 2010, \pasp, \href
  {https://ui.adsabs.harvard.edu/abs/2010PASP..122..991D} {122, 991}

\bibitem[\protect\citeauthoryear{{Di Criscienzo}, {Caputo}, {Marconi}  \&
  {Cassisi}}{{Di Criscienzo} et~al.}{2007}]{Di_Criscienzo2007}
{Di Criscienzo} M.,  {Caputo} F.,  {Marconi} M.,   {Cassisi} S.,  2007, \aap,
  \href {https://ui.adsabs.harvard.edu/abs/2007A&A...471..893D} {471, 893}

\bibitem[\protect\citeauthoryear{{Dotter} et~al.,}{{Dotter}
  et~al.}{2010}]{Dotter2010}
{Dotter} A.,  et~al., 2010, \apj, \href
  {https://ui.adsabs.harvard.edu/abs/2010ApJ...708..698D} {708, 698}

\bibitem[\protect\citeauthoryear{{Dworetsky}}{{Dworetsky}}{1983}]{Dworetsky1983}
{Dworetsky} M.~M.,  1983, \mnras, \href
  {https://ui.adsabs.harvard.edu/abs/1983MNRAS.203..917D} {203, 917}

\bibitem[\protect\citeauthoryear{{Gaia Collaboration} et~al.,}{{Gaia
  Collaboration} et~al.}{2018}]{Gaia2018}
{Gaia Collaboration} et~al., 2018, \mn@doi [\aap]
  {10.1051/0004-6361/201833051}, \href
  {https://ui.adsabs.harvard.edu/abs/2018A&A...616A...1G} {616, A1}

\bibitem[\protect\citeauthoryear{{Hachenberg}}{{Hachenberg}}{1939}]{Hachenberg1939}
{Hachenberg} O.,  1939, \zap, \href
  {https://ui.adsabs.harvard.edu/abs/1939ZA.....18...49H} {18, 49}

\bibitem[\protect\citeauthoryear{{Harris}}{{Harris}}{1996}]{Harris1996}
{Harris} W.~E.,  1996, \aj, \href
  {http://adsabs.harvard.edu/abs/1996AJ....112.1487H} {112, 1487}

\bibitem[\protect\citeauthoryear{{Irrgang}, {Wilcox}, {Tucker}  \&
  {Schiefelbein}}{{Irrgang} et~al.}{2013}]{Irrgang2013}
{Irrgang} A.,  {Wilcox} B.,  {Tucker} E.,   {Schiefelbein} L.,  2013, \aap,
  \href {https://ui.adsabs.harvard.edu/abs/2013A&A...549A.137I} {549, A137}

\bibitem[\protect\citeauthoryear{{Jurcsik} et~al.,}{{Jurcsik}
  et~al.}{2015}]{Jurcsik2015}
{Jurcsik} J.,  et~al., 2015, \apjs, \href
  {https://ui.adsabs.harvard.edu/abs/2015ApJS..219...25J} {219, 25}

\bibitem[\protect\citeauthoryear{{Kopacki}}{{Kopacki}}{2001}]{Kopacki2001}
{Kopacki} G.,  2001, \aap, \href
  {https://ui.adsabs.harvard.edu/abs/2001A&A...369..862K} {369, 862}

\bibitem[\protect\citeauthoryear{{Kopacki}}{{Kopacki}}{2007}]{Kopacki2007}
{Kopacki} G.,  2007, \actaa, \href
  {https://ui.adsabs.harvard.edu/abs/2007AcA....57...49K} {57, 49}

\bibitem[\protect\citeauthoryear{{Kov{\'a}cs} \& {Kanbur}}{{Kov{\'a}cs} \&
  {Kanbur}}{1998}]{Kovacs1998}
{Kov{\'a}cs} G.,  {Kanbur} S.~M.,  1998, \mnras, \href
  {http://adsabs.harvard.edu/abs/1998MNRAS.295..834K} {295, 834}

\bibitem[\protect\citeauthoryear{{Kov{\'a}cs} \& {Walker}}{{Kov{\'a}cs} \&
  {Walker}}{2001}]{Kovacs2001}
{Kov{\'a}cs} G.,  {Walker} A.~R.,  2001, \aap, \href
  {http://adsabs.harvard.edu/abs/2001A%26A...374..264K} {374, 264}

\bibitem[\protect\citeauthoryear{{Kunder}, {Stetson}, {Catelan}, {Walker}  \&
  {Amigo}}{{Kunder} et~al.}{2013a}]{Kunder2013b}
{Kunder} A.,  {Stetson} P.~B.,  {Catelan} M.,  {Walker} A.~R.,   {Amigo} P.,
  2013a, \aj, \href {http://adsabs.harvard.edu/abs/2013AJ....145...33K} {145,
  33}

\bibitem[\protect\citeauthoryear{{Kunder} et~al.,}{{Kunder}
  et~al.}{2013b}]{Kunder2013a}
{Kunder} A.,  et~al., 2013b, \aj, \href
  {http://adsabs.harvard.edu/abs/2013AJ....146..119K} {146, 119}

\bibitem[\protect\citeauthoryear{{Lenz} \& {Breger}}{{Lenz} \&
  {Breger}}{2005}]{Lenz2005}
{Lenz} P.,  {Breger} M.,  2005, Communications in Asteroseismology, \href
  {https://ui.adsabs.harvard.edu/abs/2005CoAst.146...53L} {146, 53}

\bibitem[\protect\citeauthoryear{{Mar\'in}}{{Mar\'in}}{2002}]{Marin2002}
{Mar\'in} Z.,  2002, Master thesis, Universidad de Los Andes, M\'erida,
  Venezuela, \href {http://adsabs.harvard.edu/abs/2009A%26A...508..695C} {}

\bibitem[\protect\citeauthoryear{{Matsunaga} et~al.,}{{Matsunaga}
  et~al.}{2006}]{Matsunaga2006}
{Matsunaga} N.,  et~al., 2006, \mnras, \href
  {https://ui.adsabs.harvard.edu/abs/2006MNRAS.370.1979M} {370, 1979}

\bibitem[\protect\citeauthoryear{{Nemec}}{{Nemec}}{2004}]{Nemec2004}
{Nemec} J.~M.,  2004, \mn@doi [\aj] {10.1086/382903}, \href
  {https://ui.adsabs.harvard.edu/abs/2004AJ....127.2185N} {127, 2185}

\bibitem[\protect\citeauthoryear{{Nemec}, {Linnell Nemec}  \& {Lutz}}{{Nemec}
  et~al.}{1993}]{Nemec1993}
{Nemec} J.~M.,  {Linnell Nemec} A.~F.,   {Lutz} T.~E.,  1993, {SX Phe Stars as
  Distance Indicators}.
p.~145

\bibitem[\protect\citeauthoryear{{Nemec}, {Cohen}, {Ripepi}, {Derekas},
  {Moskalik}, {Sesar}, {Chadid}  \& {Bruntt}}{{Nemec} et~al.}{2013}]{Nemec2013}
{Nemec} J.~M.,  {Cohen} J.~G.,  {Ripepi} V.,  {Derekas} A.,  {Moskalik} P.,
  {Sesar} B.,  {Chadid} M.,   {Bruntt} H.,  2013, \apj, \href
  {https://ui.adsabs.harvard.edu/abs/2013ApJ...773..181N} {773, 181}

\bibitem[\protect\citeauthoryear{{Pritzl}, {Armandroff}, {Jacoby}  \& {Da
  Costa}}{{Pritzl} et~al.}{2002}]{Pritzl2002}
{Pritzl} B.~J.,  {Armandroff} T.~E.,  {Jacoby} G.~H.,   {Da Costa} G.~S.,
  2002, \aj, \href {https://ui.adsabs.harvard.edu/abs/2002AJ....124.1464P}
  {124, 1464}

\bibitem[\protect\citeauthoryear{{Rees}}{{Rees}}{1992}]{Rees1992}
{Rees} Richard~F. J.,  1992, \aj, \href
  {https://ui.adsabs.harvard.edu/abs/1992AJ....103.1573R} {103, 1573}

\bibitem[\protect\citeauthoryear{{Sandage} \& {Cacciari}}{{Sandage} \&
  {Cacciari}}{1990}]{SanCac1990}
{Sandage} A.,  {Cacciari} C.,  1990, \mn@doi [\apj] {10.1086/168417}, \href
  {https://ui.adsabs.harvard.edu/\#abs/1990ApJ...350..645S} {350, 645}

\bibitem[\protect\citeauthoryear{{Shapley}}{{Shapley}}{1918}]{Shapley1918}
{Shapley} H.,  1918, \mn@doi [\apj] {10.1086/142423}, \href
  {https://ui.adsabs.harvard.edu/abs/1918ApJ....48..154S} {48, 154}

\bibitem[\protect\citeauthoryear{{Stetson}}{{Stetson}}{2000}]{Stetson2000}
{Stetson} P.~B.,  2000, \pasp, \href
  {http://adsabs.harvard.edu/abs/2000PASP..112..925S} {112, 925}

\bibitem[\protect\citeauthoryear{{Van Albada} \& {Baker}}{{Van Albada} \&
  {Baker}}{1971}]{vanAlbada1971}
{Van Albada} T.~S.,  {Baker} N.,  1971, \apj, \href
  {http://adsabs.harvard.edu/abs/1971ApJ...169..311V} {169, 311}

\bibitem[\protect\citeauthoryear{{VandenBerg} \& {Clem}}{{VandenBerg} \&
  {Clem}}{2003}]{VandenBerg2003}
{VandenBerg} D.~A.,  {Clem} J.~L.,  2003, \mn@doi [\aj] {10.1086/376840}, \href
  {https://ui.adsabs.harvard.edu/abs/2003AJ....126..778V} {126, 778}

\bibitem[\protect\citeauthoryear{{VandenBerg}, {Bergbusch}  \&
  {Dowler}}{{VandenBerg} et~al.}{2006}]{VandenBerg2006}
{VandenBerg} D.~A.,  {Bergbusch} P.~A.,   {Dowler} P.~D.,  2006, \mn@doi
  [\apjs] {10.1086/498451}, \href
  {https://ui.adsabs.harvard.edu/abs/2006ApJS..162..375V} {162, 375}

\bibitem[\protect\citeauthoryear{{VandenBerg}, {Bergbusch}, {Ferguson}  \&
  {Edvardsson}}{{VandenBerg} et~al.}{2014}]{VandenBerg2014}
{VandenBerg} D.~A.,  {Bergbusch} P.~A.,  {Ferguson} J.~W.,   {Edvardsson} B.,
  2014, \apj, \href {http://adsabs.harvard.edu/abs/2014ApJ...794...72V} {794,
  72}

\bibitem[\protect\citeauthoryear{{VandenBerg}, {Denissenkov}  \&
  {Catelan}}{{VandenBerg} et~al.}{2016}]{VandenBerg2016}
{VandenBerg} D.~A.,  {Denissenkov} P.~A.,   {Catelan} M.,  2016, \apj, \href
  {https://ui.adsabs.harvard.edu/abs/2016ApJ...827....2V} {827, 2}

\bibitem[\protect\citeauthoryear{{Walker}}{{Walker}}{1955}]{Walker1955}
{Walker} M.~F.,  1955, \aj, \href
  {https://ui.adsabs.harvard.edu/abs/1955AJ.....60..197W} {60, 197}

\bibitem[\protect\citeauthoryear{{Yepez}, {Arellano Ferro}, {Muneer}  \&
  {Giridhar}}{{Yepez} et~al.}{2018}]{Yepez2018}
{Yepez} M.~A.,  {Arellano Ferro} A.,  {Muneer} S.,   {Giridhar} S.,  2018,
  \rmxaa, \href {https://ui.adsabs.harvard.edu/abs/2018RMxAA..54...15Y} {54,
  15}

\bibitem[\protect\citeauthoryear{Zhang, Ramakrishnan  \& Livny}{Zhang
  et~al.}{1996}]{Zhang1996}
Zhang T.,  Ramakrishnan R.,   Livny M.,  1996, SIGMOD Rec., 25, 103

\bibitem[\protect\citeauthoryear{{Zinn} \& {West}}{{Zinn} \&
  {West}}{1984}]{Zinn1984}
{Zinn} R.,  {West} M.~J.,  1984, \apjs, \href
  {http://adsabs.harvard.edu/abs/1984ApJS...55...45Z} {55, 45}

\makeatother
\end{thebibliography}

\appendix
\label{appendix}

\section{Individual comments on variable stars}
\label{IndStars}

\subsection{RR Lyrae stars}

V29. The magnitude-weighted mean, colour and magnitude place this star  below the HB. Its distance derived from the Fourier decomposition (see $\S$ \ref{FOURIER}) indicates that V29 is behind the cluster. The proper motion membership analysis also suggests that V29 is not a cluster member (see Fig. \ref{PMM92}). Hence, it was not used in the calculation of the mean physical parameters.

V31. This is a RRc star sitting slightly above the HB. Our light curve at the coordinates given by the CVSGC includes two $Gaia$ sources, which unfortunately do not possess $Gaia$ colour $G_{BP}-G_{RP}$ to correct our photometry for neighbour contamination. The blinking of our residual images collection clearly identifies the variable with the $Gaia$ source 1360405469102216576, whose detailed identification is provided in Fig. \ref{ID_CHART}. According to the star membership analysis, the star is likely a cluster member. However the proper motion of the variable $Gaia$ source is at odds  with the cluster general motion (see Fig. \ref{PMM92}), implying that the star is not a cluster member. This variable is flanked by two brighter stars and also blended with another one only slightly brighter to the West. None of these neighbours is variable, but they pollute the light curve particularly near its maximum brightness (See Fig. \ref{mosaico_RR}). 

F1. This is a newly found RRc star above the HB. It is a rather isolated star matching a $Gaia$ source, whose proper motion suggests that the star is not a cluster member. Its position on the CMD about half a magnitude above the HB is also anomalous, suggesting the star is in front of the cluster.

\subsection{SX Phe stars}
\label{sec:SX}

The SX Phe stars known in M92 were discovered by \citet{Kopacki2001} (V33 and V34), \citet{Kopacki2007} (V35, V36, V37, V38 and V39) and V41 in the present work. They are all faint stars often in crowded regions or with near neighbours of similar brightness. In our photometry we have not always been able to resolve these stars and, as a consequence, a few of them appear in peculiar places in the CMD, which do not necessarily imply that the star is not a cluster member. We have searched for $Gaia$-DR2 sources and their proper motions for these stars with the aim of shedding light on its membership, and have made an effort to accurately identify their positions in Fig \ref{ID_CHART}.

The stars V33, V34, V35, V36 and V38 are found away from the Blue Stragglers region of the CMD, where they would be naturally expected if they were cluster members. This may be due to contamination of our photometry by neighbouring stars or due to the fact that the star is indeed not a cluster member. In the following paragraphs we discuss each particular case.

V33. Within the PSF of our light curve there are two $Gaia$ sources (see Fig. \ref{ID_CHART}), hence the mean magnitude is spurious and about 0.8 magnitude brighter than the light curve reported by \citet{Kopacki2007}. Blinking our image collection we associate the variations with the star to the north of the pair which is the $Gaia$ source 1360405675256740224. Unfortunately we were not able to correct the magnitude of the star and its colour since the neighbouring $Gaia$ source does not have $Gaia$ photometric indices. The proper motion of V33 (dark blue in Fig. \ref{PMM92}) seems consistent with the cluster motion, hence the star is likely a cluster member

V34. This star was reported by \citet{Kopacki2007} as a double mode $V\sim 17.1$ mag SX Phe, and the location of the star reported by him is flanked by two rather brighter stars. In our images we were unable to find the star at that precise position. However, we have found variability of the same period as the one reported by \citet{Kopacki2007}, in the west star of the pair (see the id chart in Fig. \ref{ID_CHART}), which is a $V\sim 15.4 $ mag star. The dispersion of our light curve is much smaller than that of \citet{Kopacki2007} (see Fig. \ref{mosaicoSX}), and we fear that its light curve may be contaminated by the neighboring star to the east. While it is true that the time-base of our data may not be long enough, we were unable to find a second frequency in the power spectrum after prewhitening its main frequency. However, we note that our light curve does not show traces of a probable second mode. This variable corresponds to the $Gaia$ source 1360405434742361344 and its proper motion vector suggests that the star is not a cluster member as it does its peculiar position in the CMD. There is no other $Gaia$ source nearby. 

V35. The identification of this star in \citet{Kopacki2007} map contains two or probably three stars of similar magnitude. The PSF of our photometry contains at least the two brightest stars. While we have been able to detect the variations and reproduce the period found by \citet{Kopacki2007} our light curve is scattered and the mean magnitude is spurious. There is no $Gaia$ source associated to this star, hence we also lack a proper motion. We cannot tell whether the star belongs to the cluster or not. In Fig. \ref{ID_CHART} we mark the position according to our photometry.

V36. This star was reported by \citet{Kopacki2007} as a SX Phe with $V \sim 17.88$ mag and located in the blue stragglers region of the CDM. As with for V34, our photometry contains the flux of at least two stars of similar brightness, hence, the contamination shifts the position of the star near the RGB in the CMD (Fig. \ref{CMD_M92}). Due to this, its light curve, shown in Fig. \ref{mosaicoSX}, shows some scattering. The variable star is most likely the one identified in Fig. \ref{ID_CHART} and corresponds to the $Gaia$ source 1360405469102024064 whose $G_B$ magnitude is 15.233. i.e very similar to our $V$ magnitude. According to its proper motion (Fig. \ref{PMM92}) the star is not a cluster member. There is no other $Gaia$ source nearby.

V37. This star falls in the edges of the Blue Straggler region in the CMD. The star is located to the north of a very bright neighbour, probably producing part of the scatter in the light curve. There are two $Gaia$ sources within our PSF (see Fig. \ref{ID_CHART}). Like in the case of V33, the neighbouring $Gaia$ source does not have photometric indices that may enable us to correct the star position on the CMD. In Fig. \ref{PMM92} we have plotted the proper motion of both stars, which makes clear that the star identified in Fig. \ref{ID_CHART} is a cluster member.

V38. The proper motion of this SX Phe star suggests that the star is a cluster member. However its position on the CMD in the sub-giant region is peculiar. In Fig. \ref{ID_CHART}, is evident that our photometry includes a few rather faint stars which may not affect much its magnitude, but seems to redden its photometry.

V39. This faint SX Phe star shows a very low amplitude variation, consistent with the flux-curve reported by \citet{Kopacki2007}. In the CMD it sits at the core of the blue stragglers region and its proper motion is consistent with that of the cluster. The star is most likely a cluster member.

\subsection{Anomalous Cepheid and BL Her stars}
\label{AC}

V7. It was suggested by \citet{Matsunaga2006} and \citet{Di_Criscienzo2007} that this star is an anomalous cepheid based on the fact that it is too bright to be a BL Her star. Given the mean cluster distance found from the RR Lyrae stars and reddening, the absolute magnitude of the star is $M_V = -0.45$ mag. This and the pulsating period (1.0611 d), log P = 0.025, indicate that the star follows the P-L relation of anomalous cepheids \citep[their fig. 6]{Pritzl2002}, which confirms the idea of its AC nature.

V30. \citet{Kopacki2001} discovered the variability of this star and classified it as RRab. According to his identification chart, the star is virtually at the cluster center and surrounded by three brighter stars. We inspected this area by blinking our collection of residual images and noted clear variations at the coordinates $\alpha = 17^{h}17'07.48'', \delta = +43^{\circ} 08' 06.0''$ which correspond to the $Gaia$ source 1360405469102036480. The $Gaia$ magnitude of this source is $G=13.65$, i.e. similar to the $V$ magnitude-weighted mean of our light curve (Fig. \ref{mosaico_AC}). A detailed identification is given in Fig. \ref{ID_CHART}. Although the light curve of V30 shows contamination near the maximum, most likely from the brighter neighbours, it is well defined. This star is more than a magnitude above the HB to be a RR Lyrae member of M92. However, the $Gaia$ proper motion suggests that the star is not a cluster member (Fig. \ref{PMM92}), and is rather moving a bit faster.

While at first glance its $V$ light curve and period (0.528331 d) resembles in fact a RRab star, a closer look reveals that its amplitude is too small for its period (see its odd position in Fig. \ref{figBailey}, ruling it out as an RR Lyrae star) and the descending branch is shorter than in RRab stars.

If it was a cluster member, it would have an absolute magnitude of $M_V=-0.71$ mag, which along with its period, log P = $-0.277$, places the star on the P-L relation of anomalous cepheid stars pulsating in the first overtone \citep[their fig. 6]{Pritzl2002}. However, since there is some evidence that it is not a member, we consider it to be a probable BL Her star.

% Don't change these lines
\bsp	% typesetting comment
\label{lastpage}
\end{document}